\documentclass[twocolumn,showpacs,preprintnumbers,amsmath,prl,amssymb,superscriptaddress]{revtex4-1}

\usepackage[export]{adjustbox}
\usepackage{amsfonts}
\usepackage{amsmath}
\usepackage[makeroom]{cancel}
\usepackage[english]{babel}
\usepackage[T1]{fontenc}
\usepackage{txfontsb}
\usepackage{times}
\usepackage{mathrsfs}
\usepackage{graphicx}
\usepackage{dcolumn}
\usepackage{bm}
\usepackage{wasysym}
\usepackage[colorlinks,bookmarks=true,citecolor=blue,linkcolor=red,urlcolor=blue]{hyperref}
\usepackage[tight, FIGTOPCAP, hang, raggedright, nooneline]{subfigure}
\usepackage{hyperref}
\usepackage{xcolor}
\usepackage{epsfig}
\usepackage{amssymb}
\usepackage{lipsum}

\newcommand{\be}{\begin{equation}}
\newcommand{\ee}{\end{equation}}

\begin{document}

\title{Topological electrostatics}

\author{B.~Dou\c{c}ot}
\affiliation{LPTHE, CNRS and Sorbonne Universit\'e, 75252 Paris Cedex 05, France}
\author{R.~Moessner}
\affiliation{Max-Planck-Institut f\"ur Physik komplexer Systeme, 01187 Dresden, Germany}
\author{D.~L.~Kovrizhin}
\affiliation{LPTM, CY Cergy Paris Universite, UMR CNRS 8089, Pontoise 95032 Cergy-Pontoise Cedex, France}

\date{\today}

\begin{abstract}
We present a theory of optimal topological textures in nonlinear sigma-models with degrees of freedom living in the Grassmannian $\mathrm{Gr}(M,N)$ manifold. These textures describe skyrmion lattices of $N$-component fermions in a quantising magnetic field,  relevant to the physics of graphene, bilayer and other multicomponent quantum Hall systems near integer filling factors $\nu>1$. We derive analytically the optimality condition, minimizing topological charge density fluctuations, for a general Grassmannian sigma model $\mathrm{Gr}(M,N)$ on a sphere and a torus, together with counting arguments which show that for any filling factor and number of components there is a critical value of topological charge $d_c$ above which there are no optimal textures. Below $d_c$ a solution of the optimality condition on a torus is unique, while in the case of a sphere one has, in general, a continuum of solutions corresponding to new {\it non-Goldstone} zero modes, whose degeneracy is not lifted (via a order from disorder mechanism) by any fermion interactions depending only on the distance on a sphere. We supplement our general theoretical considerations with the exact analytical results for the case of $\mathrm{Gr}(2,4)$, appropriate for recent experiments in graphene.
\end{abstract}

\maketitle

\textit{Introduction}. The theory of non-linear sigma-models \cite{rajaraman82} is a venerable subject with applications ranging from high-energy physics and black holes to soft-matter and solid state physics. In the latter setting they provide effective descriptions of quantum Hall ferromagnets and their topological excitations, such as skyrmions \cite{sondhi,girvin99}. On the other hand, these models with remarkable mathematical structures offer important insights into non-linear phenomena and geometry, and have been a topic of extensive mathematical research. This paper builds on the latter, leveraging the uncovered beautiful mathematical structures to address the problem of finding optimal topological textures in Grassmannian sigma-models. The questions related to geometry of Grassmannian manifolds have recently attracted a lot of attention in such diverse fields as string theory \cite{AHamed}, statistical mechanics \cite{Huang, Galashin}, and machine learning \cite{Zhang}. The results presented in this Letter may be also relevant to the mathematical questions of stability of vector bundles, and finding conditions for flat metrics. In our setting the optimal textures exhibit an almost flat (up to exponentially small terms) topological (and hence electric) charge density on a torus, thereby minimising the Coulomb interaction.  The charge density also also plays the role of an {\it emergent effective} magnetic field, and as such is central to the problem of quantisation of these models.

Grassmannian sigma-models provide long-wavelength description of quantum Hall ferromagnets \cite{sondhi,moon} hosting multicomponent fermions at filling factors $\nu>1$. These systems are expected to be realised, for-example, in multi-layer quantum Hall systems (with an approximate spin-layer degeneracy), and in spin-valley degenerate systems with a notable example of graphene. Here, the spin and valley rotate under approximate $\mathrm{SU(4)}$ transformation, 
see \cite{Goerbig06,Young12}, where recent experiments found evidence for skyrmion crystals away from integer filling factors \cite{Young20}. These systems have also been recently studied using exact diagonalisation \cite{Jolicoeur}. However, because of the large number of degrees of freedom, these studies are limited to small number of electrons. 

Here, we present a general method of finding ground state configurations of multicomponent fermions in the lowest Landau level at integer $\nu>1$ in the nonlinear sigma-model description and in presence of Coulomb interactions whose role is to minimise topological charge density fluctuations. While we do not take into account possible anisotropies relevant to real experimental systems, our work can be used as the starting point for more quantitative calculations \cite{Kovrizhin}.

\textit{Outline}. We consider Grassmannian sigma-models with the degrees of freedom defined on the manifold $\mathrm{Gr}(M,N)$ with $M$ being the filling factor, and $N$ internal states arising from degrees of freedom such as spin, valley, layer, etc. We wish to find the textures with smallest topological charge density fluctuations. Mathematically, the topological textures minimizing the energy of a pure sigma-model, without taking into account the interactions, correspond to holomorphic maps $w(z)$ from a base manifold $\mathcal{M}$, corresponding to physical space (a sphere or a torus in our case), to a Grassmannian manifold $\mathrm{Gr}(M,N)$ according to the  Bogomol'nyi–Prasad–Sommerfield bound \cite{rajaraman82}. For our purposes it is useful to reformulate the problem as an equivalent one in terms of the classification of holomorphic vector bundles over the base manifold $\mathcal{M}$, which allows one to apply the machinery of algebraic geometry. 

In this representation the maps are defined by a choice of a rank $M$ vector bundle $\mathcal{V}$ over the manifold $\mathcal{M}$ together with the choice of $N$ global holomorphic sections of $\mathcal{V}$, modulo automorphisms of $\mathcal{V}$. The latter correspond to gauge transformations, and play an important role in counting degrees of freedom, as we will show below. The sections generate the fiber of $\mathcal{V}$ over each point of $\mathcal{M}$. After choosing a basis $\sigma_1,...,\sigma_{D}$ of global sections of $\mathcal{V}$, a texture is encoded by an $N\times D$ matrix $A$. An automorphism of $\mathcal{V}$ acts on $A$ by right multiplication $A \rightarrow A\Lambda$, where $\Lambda$ is a  $D\times D$ matrix. Physical global
$\mathrm{SU(N)}$ transformations $g$, which commute with automorphisms, act by left multiplication $A \rightarrow gA$. 

Let us recall the definition of topological charge associated to a $\mathrm{Gr}(M,N)$ texture, where we wish to emphasize its geometric nature, also see Supp. Mat. On  $\mathrm{\mathbb{C}P^{\tilde{N}-1}}$, we have a natural K\"ahler metric (Fubini-Study metric), whose associated 2-form can be interpreted as
the curvature form, or Berry curvature, of a line bundle $\mathcal{O}(1)$ over $\mathrm{\mathbb{C}P^{\tilde{N}-1}}$. This bundle is the dual of the tautological line bundle over $\mathrm{\mathbb{C}P^{\tilde{N}-1}}$, which attaches its representative vector at every point. The topological charge density, associated to the texture described by the $N\times M$ matrix $w(z)$,  is obtained as the the pullback  of the natural curvature form on $\mathrm{\mathbb{C}P^{\tilde{N}-1}}$ under the composed map $i_{\mathcal{P}} w$. The latter is  the Pl\"ucker embedding from  $\mathrm{Gr}(M,N)$ to $\mathrm{\mathbb{C}P^{\tilde{N}-1}}$, with $\tilde{N}=N!/M!(N-M)!$ This can be used to associate to any $\mathrm{Gr}(M,N)$ texture a  $\mathrm{\mathbb{C}P^{\tilde{N}-1}}$ texture. In doing so, the associated bundle
over  $\mathcal{M}$ is the determinant bundle of $\mathcal{V}$, which is a rank 1 bundle (line bundle).
If the $\mathrm{Gr}(M,N)$ texture is defined by sections $s^{(1)},...,s^{(N)}$ of $\mathcal{V}$, the associated $\mathrm{\mathbb{C}P^{\tilde{N}-1}}$ texture 
is defined by sections $s^{(i_1)}\wedge...\wedge s^{(i_M)}$, with $1 \leq i_1 < ... < i_M \leq N$. If a texture is given by an $A$
matrix, the associated texture $i_{\mathcal{P}} f$ is encoded in a new matrix, denoted by $\tilde{A}=\mathcal{M}(A)B$. 
Here, $\mathcal{M}(A)$ is an $\tilde{N}$ by $\tilde{D}$ matrix obtained by taking rank $M$ minor determinants of $A$,
$\tilde{D}=D!/M!(D-M)!$, and $B$ is an $\tilde{D}$ by $\tilde{d}$ matrix, expressing the wedge products  
$\sigma_{i_1}\wedge...\wedge \sigma_{i_M}$ in a basis $\tau_1,...,\tau_{\tilde{d}}$ of global sections
of the determinant bundle of $\mathcal{V}$. 


We wish to minimise the Coulomb energy of these holomorphic textures, physically we want to find the textures which correspond to a system of interacting electrons in the lowest Landau level, and we find that the optimality condition is described by a set of nonlinear equations $\tilde{A}^{\dagger}\tilde{A}=I$,
where $I$ is the identity matrix. This is motivated by the following considerations. On the sphere $\mathcal{M}=S^{2}=\mathrm{\mathbb{C}P(1)}$,
this condition is exactly equivalent to having a constant  topological charge density. This is also consistent with our previous results on a torus \cite{Kovrizhin}. 
There, we have shown that residual spatial modulations of the topological charge density decrease exponentially in the large $N=d$ limit ($d$ being the total topological charge).  Mathematically, this corresponds to the classical limit for geometric quantization on the torus and this behaviour is just
a special case of the theory of the Bergman kernel asymptotics developed in the 90's by Tian, Yau, Zelditch, Catlin, Lu \cite{Tian,Zelditch}, and applied to
quantum Hall physics in particular by S.~Klevtsov \cite{Klevtsov}.
In practice, the $\tilde{A}^{\dagger}\tilde{A}=I$ condition can be formulated as $\mathcal{B}^{\dagger}M(A)\mathcal{B}=I$, 
where $M(A)=\mathcal{M}(A)^{\dagger}\mathcal{M}(A)$ is a $\tilde{D}$ by $\tilde{D}$ square matrix, whose entries are given by $M$-dimensional determinants whose elements are also elements of $A^{\dagger}A$, so they are invariant under global $\mathrm{SU(N)}$ transformations. We find that in general, the optimality condition has solutions on a sphere and a torus for a maximal value $d_c$ given by $N-1$ and $N$, respectively.

\textit{Grassmannian holomorphic textures on a sphere}. In order to construct optimal topological textures over
a sphere we will use the key mathematical result given by Grothendieck's theorem \cite{Grothendieck} (apparently this result has been
derived several times in the previous century, see e.g.~\cite{Birkhoff}), which states that any 
rank $M$ vector bundle on the sphere splits as a direct sum of line bundles. We need an explicit description of line bundles on
$S^{2}=\mathbb{C}P(1)$, which have global sections. They are the $\mathcal{O}(d)$ bundles, where $d$ is a positive integer equal to the
topological charge. The space of global holomorphic sections of $\mathcal{O}(d)$ on $S^{2}$ is
realized by polynomials in $z$ with maximal degree equal to $d$, so its dimension is equal to $d+1$.
Physically, this space is a realization of the Hilbert space of a quantum spin $S=d/2$. Any rank $M$ vector bundle on $S^{2}$ is of the form 
$\mathcal{V}=\mathcal{O}(d_1) \oplus \mathcal{O}(d_2)... \oplus \mathcal{O}(d_M)$,
with $d_1 + d_2 + ... + d_M = d$ the total topological charge.

\begin{figure}[t]
\epsfig{file=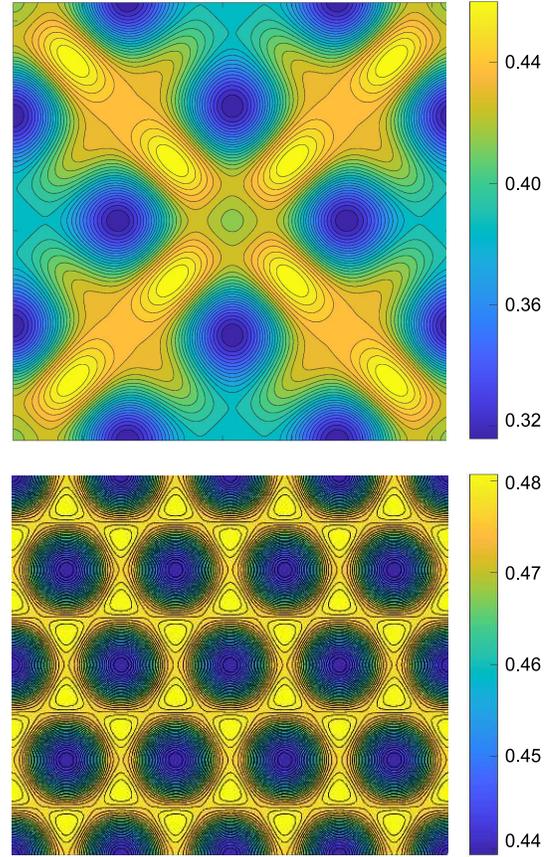,width=7.5cm}
\caption{(Color online). Topological charge density corresponding to Skyrmion textures minimizing 
charge fluctuations. Top panel: unit cell of a square lattice, obtained by numerical minimization
in the basis of theta-functions (note that the square lattice does not satisfy the optimality conditions). 
Bottom panel: unit cell of a triangular lattice corresponding to optimal texture discussed in the text.
Note the difference in scale of the topological charge fluctuations.}
\end{figure}

Let us look at specific example of the Grassmannian sigma-model $\mathrm{Gr}(M,N)$ on a sphere, where we consider the case of $M=2$.
While perhaps not the most physical example, this highlights the structure of the problem. To this end we also show a different way of
finding optimal textures in this case. A general holomorphic solution of the non-linear sigma model can be written as an $N\times M$ matrix
$w(z)$ which in case of a topological charge $d=2,\ d_1=1,\ d_2=1$  and $M=2$ can be written as $w(z)=(\mathbf{a}_1+\mathbf{a}_2 z, \mathbf{b}_1+\mathbf{b}_2 z)$, where $\mathbf{a}_i$ and $\mathbf{b}_i$ are $N$-dimensional column vectors. The topological charge density is invariant with respect to an arbitrary global $\mathrm{SU(N)}$ transformation acting on the left of $w(z)$ and a local gauge transformation given by a $M\times M$ matrix whose matrix elements are given by analytic functions, acting on the right. 

The matrix $w^{\dagger}w$ is invariant under global $\mathrm{SU(N)}$ transformations, and we can use the coefficients in the expansion of this matrix in powers of $z$ as gauge-independent parameters defining a given topological texture. For $M=2$ and $d=2$ we have $w^{\dagger}(z)w(z)=A+B|z|^2+C z+C^{\dagger}z^*$, where $A,B$ are Hermitian $2\times 2$ matrices formed from the overlaps of vectors such as $\langle \mathbf{a}_i | \mathbf{a}_j\rangle$. Using Gramm-Schmidt orthogonalization  we  find an invertible matrix $R$ such that $R^{\dagger}B R=I$, and write $A$ in diagonal form $\Lambda=U^{\dagger}(R^{\dagger}A R)U$ using an unitary transformation $U$. This yields in total six real equations on the matrix elements of $A$ and $B$, and together with the optimality conditions for the topological charge density $\mathrm{det}(w^{\dagger}w)=(1+|z|^2)^d$ with $d=2$ allows one to find the optimal texture:
\begin{equation}
w^{\dagger}(z)w(z)=\left(\begin{array}{cc}
\langle a_1|a_1\rangle+|z|^2  & \langle a_2|b_1\rangle z^{*} \\
\langle b_1|a_2\rangle  z & \langle a_1|a_1\rangle^{-1}+|z|^2 \end{array}\right).
\end{equation}
The  additional constraint $|\langle a_2| b_1\rangle|^2=\langle a_1|a_1\rangle^{-1}(\langle a_1|a_1\rangle-1)^2$ fixes the amplitude of the coefficient $\langle a_2| b_1\rangle$, whose phase can be further removed by a rotation in the complex $z$-plane. That leaves a single real parameter $\langle a_1|a_1\rangle$ characterising the texture --- a non-Goldstone zero-mode! The possibility of these new zero-modes on a sphere is supported by our general counting arguments. It is worth noting that standard counting arguments have to be modified by taking into account automorphisms, as explained in the Supp.~Mat, also see \cite{Lomadze}.

\textit{Grassmannian textures on the torus}. In contrast to the sphere, it is known that indecomposable vector bundles exist on a torus \cite{Atiyah57}. For us these are more interesting than decomposable ones because their number of automorphisms is typically lower for a given topological charge. This suggests a larger phase space to construct optimal textures. We need to describe these explicitly, in particular to find the basis of
their spaces of holomorphic sections and automorphisms.

First, let us address the question of constructing the basis of holomorphic sections over a torus. While this question has been studied in the mathematical literature \cite{Iena09, Polishchuk} it is worth presenting an explicit construction here. Consider a torus $\mathrm{T}=\mathbb{C}/\mathbb{Z}\gamma_1\oplus\mathbb{Z}\gamma_2$ defined by two complex translation vectors $\gamma=n_1\gamma_1+n_2\gamma_2$ with $n_{1,2}\in \mathbb{Z}$. The theta-functions of a given type $(a_\gamma,b_\gamma)$ are defined by the following condition $\theta(z+\gamma)=e^{a_\gamma z+b_\gamma}\theta(z)$, where $z$ is the complex coordinate on the torus. We can use these to construct an infinite-rank vector bundle over the torus in the following way. Let sections of this bundle be represented by infinite row-vectors $s(z)=(s_0(z),s_1(z),\ldots)$. After introducing orthonormal basis vectors in this space $\hat{l}_m=(0,\ldots,1_m,\ldots)$ a section can be written as
\begin{equation}
s^{(n)}(z)=\sum_{q=0}^{\infty}\frac{1}{q!}\frac{d^q \theta(z)}{dz^q}\hat{l}_{n+q},
\end{equation}
for $n\ge0$. These sections have the following ``periodicity'' 
\begin{equation}\label{eq1}
s(z+\gamma)=e^{a_\gamma z+b_\gamma}s(z)e^{a_\gamma J},
\end{equation}
where $J$ is a square matrix with $J_{nn+1}=1$ for all $n\ge0$ and zero otherwise. This construction also holds in a finite-dimensional setting, where we set $\hat{l}_n=0$ for $n\ge M$ in which case $J$ is an $M\times M$ matrix. If the degree of the theta-functions is $d'$, the space of sections of a degree $d=M d'$ vector bundle is generated by $s^{(n)}(z)$ with $0\le n\le M-1$ and its dimension equals $d$. The sections satisfying this property are holomorphic sections of a rank $M$ indecomposable vector bundle over the torus \cite{Atiyah57, Polishchuk}. Note that this construction addresses the case when $d$ is a multiple of $M$. The general construction of indecomposable vector bundles is further explained in the Suppl. Mat.

The automorphisms acting on these sections on the right are  $M\times M$ invertible matrices $\Phi(z)$ with coefficients analytic functions of $z$. The equation $(\ref{eq1})$ should be satisfied also for these bundles which constrains the matrices $\Phi(z)$ by the condition $e^{a_\gamma J}\Phi(z+\gamma)=\Phi(z)e^{a_\gamma J}$. Using this condition, we find that 
when $d$ is a multiple of $M$,
the matrix of automorphisms $\Phi$ is given by the constant matrix with main diagonal and all upper diagonals filled with constants $\lambda_0,\ldots,\lambda_{M-1}$ correspondingly. This gives us in total $M-1$ non-trivial automorphisms. Note that there are no non-trivial automorphisms in the opposite case when $d$ and $M$ are relatively prime numbers.

Let us apply the construction given above in the simplest non-trivial case of $M=2$. The basis of degree $d'$ theta-functions with the given type characterised by two complex parameters $(a_\gamma,b_\gamma)$ will be denoted by $\{\theta_{0}(z),\theta_1(z),\ldots\theta_{d'-1}(z)\}$, see the definitions of the theta-functions below. The corresponding basis of sections of rank $2$ vector bundle $\{\sigma_0(z),\ldots\sigma_{d-1}(z)\}$ is $\sigma_j(z)=(\theta_j(z),\theta'_j(z))$ and $\sigma_{j+d'}(z)=(0,\theta_j(z))$ for $j=0,\ldots d'-1$. With this basis, any holomorphic texture can be constructed by choosing $N$ sections $s^{(1)}(z),\ldots s^{N}(z)$, which can be written in terms of the expansion in the $\sigma$-basis, namely $s^{i}(z)=\sum_1^d A_{ij}\sigma_j(z)$, with $A$ an $N\times d$ matrix of complex coefficients. Application of the Pl\"ucker map results in a corresponding $\mathrm{\mathbb{C}P^{\tilde N-1}}$ texture over the torus. For any pair of sections $(s^{(1)},s^{(2)})$ of $\mathcal V$ we can construct a section $s^{(1)}\wedge s^{(2)}$ of a determinantal bundle $\mathcal Det\ V$, which satisfies 
\begin{equation}
s^{(1)}(z+\gamma)\wedge s^{(2)}(z+\gamma)=e^{2(a_\gamma z+b_\gamma)}s^{(1)}(z)\wedge s^{(2)}(z),
\end{equation} 
which defines theta-functions $\tilde\theta(z)$ with charge $d=2d'$ and type $(\tilde a_\gamma,\tilde b_\gamma)=2(a_\gamma,b_\gamma)$. Taking a $d$-dimensional orthonormal basis $\tilde\theta_j(z)$ of these theta-functions, the Pl\"ucker map is encoded in the coefficients $\mathcal{B}_{j_1j_2,j}$ of the expansion
\begin{equation}
\sigma_{j_1}(z)\wedge\sigma_{j_2}(z)=\sum_{j=0}^{d-1} \mathcal{B}_{j_1j_2,j}\tilde\theta_j(z),
\end{equation}
with $0\le j_1< j_2\le d-1$. These coefficients are obtained by expanding the products of theta-functions $\theta_j(z)$ and their derivatives in the basis of $\tilde\theta_j(z)$, see below. 

\textit{Theta functions and expansion coefficients}. Here we focus on the case of even $d$.  We need to define theta-functions of degree $d$ over the $(\gamma_1,\gamma_2)$ torus. Let us choose $\gamma_1=\pi$ and $\gamma_2=\pi\tau$ with $\mathrm{Im}\tau>0$. It is convenient to introduce $q=e^{i\pi\tau}$ and we fix the type of the theta-functions by choosing $a_{\gamma_1}=0$, $a_{\gamma_2}=-2id$, and $b_{\gamma_{1,2}}=0$. The theta-functions, which provide an orthonormal basis, are 
\begin{equation}
\theta_p(z)=\sum_{n\in\mathbb{Z}}q^{d(n-p/d)(n-p/d-1)}e^{2 i (n d-p)z},
\end{equation}
for $p=0,\ldots,d-1$. Similarly one can define the functions $\theta_p(z)$ by changing $d$ to $d'$ in the expression above. In order to calculate the coefficients of the matrix $\mathcal{B}$  we need the expansion coefficients of the products of $\theta(z)$ and their derivatives on the basis of $\tilde\theta(z)$:
\begin{align}
&\theta_{m}(z)\theta_{n}(z)=c_{n-m}\tilde\theta_{n+m}(z)+c_{n-m+d'}\tilde\theta_{n+m-d'}(z),\\[4pt]
&\theta_{m}(z)\theta'_{n}(z)-\theta'_{m}(z)\theta_{n}(z)=-2 i d[r_{n-m}\tilde\theta_{n+m}(z)+r_{n-m+d'}\tilde\theta_{n+m-d'}(z)],\nonumber
\end{align}
where $c_k=\sum_{m\in\mathbb{Z}}q^{d(m+k/d)^2}$ and $r_k=\sum_{m\in \mathbb{Z}}(m+k/d)q^{d(m+k/d)^2}$.

Given the $N\times d$ matrix $A$ defining the $\mathrm{Gr}(2,N)$ texture, the corresponding texture in $\mathrm{\mathbb{C}P^{\tilde N-1}}$ under the Pl\"ucker map is defined by the $\tilde N\times d$ matrix $\tilde A$:
\begin{equation}
\tilde{A}_{i_1i_2,j}=\sum_{0\le j_1<j_2\le d-1}\begin{vmatrix}A_{i_1 j_1} & A_{i_1 j_2}\\ A_{i_2 j_1} & A_{i_2 j_2}\end{vmatrix} \mathcal{B}_{j_1j_2,j}.
\end{equation}
This allows one to expand the sections of $\mathcal{D}et\ \mathcal{V}$ on the basis of $\tilde\theta_j(z)$ with the coefficients given by the matrix $\tilde A$, or explicitly $s^{(i_1)}(z)\wedge s^{(i_2)}(z)=\sum_{j=0}^{d-1}\tilde A_{i_1 i_2,j}\tilde\theta_j(z)$. From this, one can read-off the optimality condition for a Grassmannian holomorphic texture on a torus, that is given by $\tilde A^{\dagger}\tilde A=I$, provided that the $\tilde\theta_j(z)$ are orthonormal. This condition can be rewritten in an explicitly $\mathrm{SU(N)}$ invariant form in terms of the coefficient of matrix $A^{\dagger}A$ as $\mathcal B^{\dagger}M(A)\mathcal{B}=I$, where
\begin{equation}
M(A)_{j_1j_2,j'_1j'_2}=\begin{vmatrix}(A^\dagger A)_{j_1 j'_1} & (A^\dagger A)_{j_1 j'_2} \\ (A^\dagger A)_{j_2 j'_1} & (A^\dagger A)_{j_2 j'_2}\end{vmatrix}.
\end{equation}
As in the case of a sphere the $\mathrm{SU(N)}$ invariant matrix elements of $A^{\dagger}A$ may be regarded as the basic degrees of freedom. We note that these degrees of freedom are fixed only up to automorphisms. 

The optimality condition presented above is one of the central results of this paper. Solving the equation $\mathcal{B}^{\dagger}M(A)\mathcal{B}=I$ in terms of the matrix elements $A^{\dagger}A$ allows one to find an optimal texture in terms 
of its expansion in theta-functions with the coefficients of this expansion given by the matrix $A$. The latter can be obtained from the matrix $A^{\dagger}A$, given that $d \leq N$, via LU-decomposition, and we present an example below.

\textit{Counting degrees of freedom on a torus}. We are now in a position to provide counting arguments for the numbers of degrees
of freedom on a torus. By the Riemann-Roch theorem \cite{GH78}, $D=d$, where  $d$ is the total topological charge. The number of independent real parameters in $\tilde{A}^{\dagger}\tilde{A}$ is $d^2$, and the number of parameters in $A^{\dagger}A$ is generically equal to $d^2$ when $d \leq N$, and to
$2 N d-N^2 < d^2$ when $d > N$. In the absence of non-trivial automorphisms, which is the case when $M$ and $d$ 
are relatively prime (e.g. odd $d$ for $M=2$), the number of constraints and the number of parameters are equal, provided $d \leq N$. 
This suggests that optimal textures correspond to a finite set of $\mathrm{SU(N)}$ orbits, and we find a solution to the optimality constraint for any odd $d$ and $M=2$. In fact, for $d=N$ this solution is unique and is given by the identity matrix $A=I$. This is clear from the fact that in the odd $d$ case one can show that $\mathcal B^\dagger \mathcal B=I$.

For even $d$ (and $M=2$), there is a 1-parameter group of non-trivial automorphisms, which act on $A^{\dagger}A$, but not on $\tilde{A}^{\dagger}\tilde{A}$ and there are in general a priori more constraints than physical parameters, so the optimality condition $\tilde A^{\dagger}A$ may not have a solution. Indeed, we find that this is the case for $\mathrm{Gr}(2,N)$ and $d=4$, where solutions exist for the skyrmion lattices with opening angles $\alpha\le\pi/3$, and it is impossible to satisfy these constraints for angles greater than $\pi/3$, even if $d \leq N$.

\textit{Grassmannian $\mathrm{Gr}(2,N)$ textures on a torus}. Here we present an explicit solution of the optimality condition in the case of $M=2$ and $d=4$. We note that the main step is to find a solution for the matrix $A^{\dagger}A$, whose size is independent of $N$. Because the matrix $A$ can then be found by LU decomposition when $N \ge d$, we focus on the case $N=4$. Let us define the matrix $\mathcal{B}$ in terms of its coefficients. We have the following coefficient structure with (in general) complex parameters which depend only on the opening angle of the lattice $\alpha$

\begin{equation}
\mathcal{B}=\begin{pmatrix}
0 & B_{1} & 0 & -B_{1}\\
B_{2} & 0 & B_{3} & 0\\
0 & B_{4} & 0 & B_{4}\\
0 & B_{4} & 0 & B_{4}\\
B_{3} & 0 & B_{2} & 0\\
0 & 0 & 0 & 0
\end{pmatrix}.
\end{equation}

The parameters $B_{i}$ can be expressed in terms of theta functions, but as it turns out we only need certain combinations
of these parameters. Let us introduce another parameter $\lambda$, given by the equation $\lambda=[-2\mathrm{Re}(B_{2}^{*}B_{3})/(|B_{2}|^{2}+|B_{3}|^{2})]^{1/2}$,
where $\lambda(\alpha)$ is a monotonic function of the opening angle of the lattice on the interval
$\alpha\in[0,\pi/3]$, with $\lambda(0)=1$ and $\lambda(\pi/3)=0$. The parameters $B_i$ are not all independent, and one can notice that
$|B_{2}-B_{3}|^{2}/ |2B_{4}|^{2}=1$. Using this relation, the general solution of the  equation $B^{\dagger}M(A)B=I$ is given by the following matrix
%
\begin{equation}
A=\left(\begin{array}{cccc}
1 & i\lambda & X & i\lambda X\\
0 & \sqrt{1-\lambda^{2}} & 0 & X\sqrt{1-\lambda^{2}}\\
0 & 0 & \frac{1}{\sqrt{2}}\frac{|B_{1}|}{|B_{4}|} & -i\lambda\frac{1}{\sqrt{2}}\frac{|B_{1}|}{|B_{4}|}\\
0 & 0 & 0 & \frac{1}{\sqrt{2}}\frac{|B_{1}|}{|B_{4}|}\sqrt{1-\lambda^{2}}
\end{array}\right),
\end{equation}
where $X$ is an arbitrary complex constant. One can see directly via the Pl\"ucker embedding that $X$ parametrises an automorphism. In other words the matrix $A$ maps to the same vector in the projective space independent of $X$. This explicitly confirms the one-parameter automorphism in the case of $M=2$, which general form for arbitrary $M$ was presented above.

To summarise, we found a general optimality condition for the topological textures in non-linear Grassmannian sigma models $\mathrm{Gr}(M,N)$ for any $M$ and $N$ and arbitrary topological charge. In case of the torus, this condition can be resolved when $M$ and $d$ are relatively prime numbers,
while in the opposite case, we find that there is a possibility of obstruction, which imposes limits on skyrmion lattice geometry. Remarkably, we find new non-Goldstone zero modes in the case of a sphere. It will be interesting to understand the physical implications of these modes, and the possible existence of other low-energy excitations. More broadly, we have uncovered a mathematically rich aspect of  topological condensed matter physics, which arises when topologically non-trivial states and their excitations are faced with `local' constraints arising from minimisation of Coulomb energies, a situation we christen topological electrostatics. 

\textit{Acknowledgements}. D.K. acknowledges partial support from Labex MME-DII (Mod\`eles Math\'ematiques et Economiques de la Dynamique, de l'Incertitude et des Interactions), ANR11-LBX-0023. B. D. gratefully acknowledges the MPIPKS for supporting his visit during the final stage of the redaction.

\appendix

\hfill\\
\noindent\makebox[\linewidth]{\resizebox{0.7\linewidth}{1pt}{$\bullet$}}\bigskip

\begin{center}
	\textbf{Supplemental Material}
\end{center}

\section{Some remarks on Grassmannian sigma-models}

\subsection{Reminder on $\mathbb{C}P(N-1)$ sigma-models}

We will consider Grassmannian sigma-models as generalisations of the $\mathbb{C}P(N-1)$ models. For the latter we have a map $r\to |\Psi(r)\rangle$, where $|\Psi(r)\rangle$ is an $N$-component spinor. The action of the $\mathbb{C}P(N-1)$ model reads:
\be
S=\int d^2 r \left(\frac{\langle\partial_\mu\Psi |\partial_\mu\Psi\rangle}{\langle\Psi |\Psi\rangle}-\frac{\langle\partial_\mu\Psi |\Psi\rangle\langle\Psi |\partial_\mu\Psi\rangle}{\langle\Psi |\Psi\rangle^2}\right).
\ee
We can also define Berry-connection $\mathcal A_\mu=\mathrm{Im}\left(\frac{\langle\Psi | \partial_\mu \Psi\rangle}{\langle\Psi |\Psi\rangle} \right)$, and topological charge density $Q(r)=\frac{1}{2\pi}(\partial_x\mathcal{A}_y-\partial_y\mathcal{A}_x)$, or explicitly
\be
Q(r)=\frac{1}{2\pi i}\left(\frac{\langle\partial_x\Psi |\partial_y\Psi\rangle}{\langle\Psi |\Psi\rangle}-\frac{\langle\partial_x\Psi |\Psi\rangle\langle\Psi |\partial_y\Psi\rangle}{\langle\Psi |\Psi\rangle^2}\right)-(x\leftrightarrow y).
\ee
Importantly, there it is possible to express these quantities in a way which allows one to generalise these constructions to the case of Grassmannian models. Let us introduce the normalised column vector $U(r)=\langle\Psi |\Psi\rangle^{-\frac{1}{2}}|\Psi\rangle$. The normalisation ensures that $U^{\dagger}U=I$. Further, we can define the orthogonal projector $P=U U^{\dagger}$ on the line generated by $|\Psi\rangle$.

Using these definitions we can rewrite the $\mathbb{C}P(N-1)$ sigma-model in terms of $U$ and $P$
\be
S=\int d^2 r [\partial_\mu U^{\dagger}\partial_\mu U-(\partial_\mu U^{\dagger})UU^{\dagger}(\partial_\mu U))],
\ee
which is equivalent to 
\be
S=\frac{1}{2}\int d^2 r \mathrm{Tr}(\partial_\mu{P}\partial_\mu{P}).
\ee
The topological charge density can be expressed in terms of $U$ or P as well. We note that in terms of $U$ the Berry phase has the form $\mathcal{A}_\mu=-i U^{\dagger}\partial_\mu U$. The topological charge density is given by $Q(r)=\frac{1}{2\pi i}[(\partial_x U^{\dagger})(\partial_y U)-(\partial_y U^{\dagger})(\partial_x U)]$, or in terms of $P$,
\be
Q(r)=\frac{1}{2\pi i}\mathrm{Tr}(P[\partial_x P,\partial_y P]).
\ee

\subsection{Generalisation to Grassmannians}

Let us consider the manifold $\mathrm{Gr}(M,N)$ of $M$-dimensional subspaces of the $N$-dimensional complex vector space $\mathbb{C}^N$. One way to parametrise such a subspace is to pick $M$ orthonormal vectors which generate it. Arranging these vectors as columns of a $N\times M$ matrix we form a matrix $U$. The orthonormality of the columns is equivalent to the constraint $U^{\dagger}U=I_M$. As in the $\mathbb{C}P(N-1)$ case, the projector on the $M$-dimensional subspace spanned by the columns of $U$ is given by the matrix $P=U U^{\dagger}$, indeed we have $P^2=P$.

For a given subspace there is a continuous manifold of orthonormal basis which generates it. This manifold is in fact $\mathrm{U(M,\mathbb{C})}$. There is therefore a redundancy in the $U$ description which can be seen as follows. A change of the orthonormal basis can be implemented by multiplying the matrix $U$ on the right by a square matrix $g\in\mathrm{U(M,\mathbb{C})}$. Such a multiplication $U\to U g$ preserves the norm provided $g^{\dagger}g=I_M$ and leaves the projector unchanged provided that $g g^{\dagger}=I_M$.

The action and the topological charge density can then be easily generalised (here $U$ is an $N\times M$ matrix):
\begin{eqnarray}
S=\int d^2 r\ \mathrm{Tr}[(\partial_\mu U^{\dagger})(\partial_\mu U)-(\partial_\mu U^{\dagger})UU^{\dagger}(\partial_\mu U))],\\
Q(r)=\frac{1}{2\pi i}\mathrm{Tr}[(\partial_x U^{\dagger})(\partial_y U)-(\partial_y U^{\dagger})(\partial_x U)],
\end{eqnarray}
or in terms of the projector
\begin{eqnarray}
S=\frac{1}{2}\int d^2 r\ \mathrm{Tr}(\partial_\mu{P}\partial_\mu{P}),\\
Q(r)=\frac{1}{2\pi i}\mathrm{Tr}(P[\partial_x P,\partial_y P]).
\end{eqnarray}
We can check directly that the action and the topological charge density are invariant under local gauge transformations $g(r)$ which act as multiplication $U\to U g(r)$.

\subsection{BPS bound}
In order to find the BPS bound which minimizes the energy of the sigma-model it is convenient to write the energy density as the square of the covariant derivative. Indeed, we have
\begin{multline}
\mathrm{Tr}[(\partial_\mu U^{\dagger})(\partial_\mu U)-(\partial_\mu U^{\dagger})UU^{\dagger}(\partial_\mu U))]\\
=\mathrm{Tr}[(\partial_\mu U^{\dagger}+U^{\dagger}(\partial_\mu U)U^{\dagger})(\partial_\mu U+U(\partial_\mu U^{\dagger})U)]\\
=\mathrm{Tr}[(\partial_\mu U^{\dagger}+(U^{\dagger}\partial_\mu U)U^{\dagger})(\partial_\mu U-U(U^{\dagger}\partial_\mu U)].
\end{multline}

We introduce the $M\times M$ matrix $\mathcal{A}_\mu=-i U^{\dagger}(\partial_\mu U)$. Because $U^{\dagger}U=I_M$, we have $\mathcal{A}_\mu=\mathcal{A}_\mu^{\dagger}$. Note that $\mathcal{A}_\mu$ transforms like a non-Abelian gauge potential under a local gauge transformation: if $U$ goes into $Ug$ (with $g^{\dagger}g=g g^{\dagger}=I_M$) then $\mathcal{A}_\mu$ goes into $g^{\dagger}\mathcal{A}_\mu g-i g^{\dagger}\partial_\mu g$. Using this gauge potential we can write the action in the following form
\be
S=\int d^2 r\ \mathrm{Tr}[(\partial_\mu U^{\dagger}+i \mathcal{A}_\mu U^{\dagger})(\partial_\mu U-i U \mathcal{A}_\mu )].
\ee
The gauge invariance of the action becomes explicit in this formulation because under $U\to Ug$ with $g^{\dagger}g=g g^{\dagger}=I_M$ we have
\begin{eqnarray}
\partial_\mu U-i U \mathcal{A}_\mu&\to (\partial_\mu U-i U \mathcal{A}_\mu)g,\\
\partial_\mu U^{\dagger}+i \mathcal{A}_\mu U^{\dagger}&\to g^\dagger(\partial_\mu U^{\dagger}+i \mathcal{A}_\mu U^{\dagger}).
\end{eqnarray}
The $\mathcal{A}_\mu$ field then can be used to define covariant derivatives: $\mathcal{D}_\mu U\equiv\partial_\mu U-i U\mathcal{A}_\mu$ and $(\mathcal{D}_\mu U)^{\dagger}\equiv\partial_\mu U^{\dagger}-i \mathcal{A}_\mu U^{\dagger}$, then
\begin{equation}
S=\int d^2 r\ \mathrm{Tr}[(\mathcal{D}_\mu U)^{\dagger}(\mathcal{D}_\mu U)].
\end{equation}
Similarly for the topological charge density we can obtain an expression in terms of covariant derivatives
\be
Q(r)=\frac{1}{2\pi i}\mathrm{Tr}[(\mathcal{D}_x U)^{\dagger}(\mathcal{D}_y U)-(\mathcal{D}_y U)^{\dagger}(\mathcal{D}_x U)],
\ee
and also $Q(r)=\frac{1}{2\pi i}\mathrm{Tr}(\partial_x\mathcal{A}_y-\partial_y\mathcal{A}_x)$, which generalises the $M=1$ case.

We can now use this to generalise the BPS bound to the Grassmannian case as follows:
\begin{multline}
\mathrm{Tr}[(\mathcal{D}_\mu U)^{\dagger}(\mathcal{D}_\mu U)]=\frac{1}{2}\mathrm{Tr}[((\mathcal{D}_x U)^{\dagger}+i (\mathcal{D}_y U)^{\dagger})
(\mathcal{D}_x U-i \mathcal{D}_y U)\\
+((\mathcal{D}_x U)^{\dagger}-i (\mathcal{D}_y U)^{\dagger})(\mathcal{D}_x U+i \mathcal{D}_y U)],
\end{multline}
which gives the inequality
\begin{multline}
\mathrm{Tr}[(\mathcal{D}_\mu U)^{\dagger}(\mathcal{D}_\mu U)]\ge |\frac{1}{2}\mathrm{Tr}[((\mathcal{D}_x U)^{\dagger}+i (\mathcal{D}_y U)^{\dagger})
(\mathcal{D}_x U-i \mathcal{D}_y U)\\
-((\mathcal{D}_x U)^{\dagger}-i (\mathcal{D}_y U)^{\dagger})(\mathcal{D}_x U+i \mathcal{D}_y U)]|.
\end{multline}
Here the last line can be written as follows
\be
\mathrm{Tr}[(\mathcal{D}_\mu U)^{\dagger}(\mathcal{D}_\mu U)]\ge 2\pi |Q(r)|,
\ee
and integrating over the whole space and introducing the integer constant $N_{top}$ for the topological charge we obtain the BPS inequality
\be
S\ge 2\pi |N_{top}|.
\ee

\subsection{Formulation in a general gauge}
We would like to obtain the expressions for the action and the topological charge density of the Grassmannian sigma model without the normalisation constraint. This formulation is convenient because we can use it to study holomorphic mappings into Grassmannian manifold. 

Let us take $V$ to be a rank $M$, $N\times M$ matrix which depends smoothly on spatial coordinates $x$ and $y$. Since the action and the topological charge density can be expressed in terms of the gauge-invariant projector $P$, we first give an expression for $P$ in terms of unnormalised matrix $V$. We have
\be
P=V(V^{\dagger}V)^{-1}V^{\dagger}.
\ee
It is clear that $P=P^{\dagger}$ and $P^2=P$, so $P$ is a projector. Then $PV=P$, so the columns of $V$ are eigenvectors of $P$ with the eigenvalue $1$. If $w$ is a vector in $\mathbb{C}^N$ chosen to be orthogonal to the columns of $V$ then $V^\dagger w=0$ and $P w=0$, and $P$ is indeed a projector onto the subspace spanned by the columns of $V$. We can now substitute the projector written in terms of $V$ into the expression for the action and after some algebra we obtain
\begin{multline}
S=\int d^2 r\ \mathrm{Tr}[\partial_\mu P \partial_\mu P]=\int d^2 r\ \mathrm{Tr}[(V^{\dagger}V)^{-1}(\partial_\mu V^{\dagger})(\partial_\mu V)\\
-(V^{\dagger}V)^{-1}(\partial_\mu V^{\dagger})V(V^{\dagger}V)^{-1}V^{\dagger}(\partial_\mu V)].
\end{multline}
Similarly, we obtain for the topological charge density
\begin{multline}
Q(r)=\frac{1}{2\pi i}\mathrm{Tr}[(V^{\dagger}V)^{-1}(\partial_x V^{\dagger})(\partial_y V)\\
-(V^{\dagger}V)^{-1}(\partial_x V^{\dagger})V(V^{\dagger}V)^{-1}V^{\dagger}(\partial_y V)]-[x\leftrightarrow y],
\end{multline}
which simplifies in the case of holomorphic $V(z)$ where we obtain
\be
Q(r)=\frac{1}{\pi}\frac{\partial}{\partial z}\frac{\partial}{\partial \bar z}\log\mathrm{Det}(V^\dagger V).
\ee

\section{$\mathrm{Gr(2,N)}$ textures on the torus associated to indecomposable rank 2 bundles. Case I.}
\subsection{Taking derivatives of theta-functions}
We fix a type $(a_\gamma,b_\gamma)$ where $(\gamma=n_1\gamma_1+n_2\gamma_2$, and $n_1,n_2$ are integers) for the theta-functions
on the torus $\mathbb{T}=\mathbb{C}/\mathbb{Z}\gamma_1\oplus\mathbb{Z}\gamma_2.$ In other words, we have the following quasi-periodicity property
\be
\theta(z+\gamma)=\exp(a_\gamma z+ b_\gamma)\theta(z).
\ee
We construct an infinite rank vector bundle over the torus using the following prescription. Let sections of this bundle be represented
by the infinite row-vectors: $s(z)\equiv (s_0(z),s_1(z),\ldots)$. We introduce an infinite basis in this space of row-vectors
\begin{eqnarray}
l_0=(1,0,0,\ldots)\\
l_1=(0,1,0,\ldots)\\
\ldots
\end{eqnarray}
together with the matrix $J$ defined by $J_{nn'}=1$ if $n'=n+1$ and $J_{nn'}=0$ otherwise $(n,n'\ge 0)$, we have $l_n J=l_{n+1}$.

Let us pick a $\theta$-function with the $(a_\gamma,b_\gamma)$ type and consider the section
\be
s(z)\equiv\sum_{q=0}^\infty \frac{1}{q!}\frac{d^{q}\theta(z)}{dz^q}l_{n+q}
\ee
for a given $n\ge 0$. The key remark is that these sections transform in a very simple way under transformations by $\gamma$
\begin{multline}
s(z+\gamma)=\sum_{q=0}^\infty \frac{1}{q!}\frac{d^{q}}{dz^q}(e^{a_\gamma z+ b_\gamma}\theta(z))l_{n+q}\\
=\sum_{q=0}^\infty\sum_{p=0}^q e^{a_\gamma z+ b_\gamma}\frac{1}{q!}\frac{q!}{p!(q-p)!}\frac{d^{q}\theta(z)}{dz^q}a_\gamma^{q-p}l_{n+q}\\
=e^{a_\gamma z+ b_\gamma}\sum_{p=0}^\infty\frac{1}{p!} \frac{d^{p}\theta(z)}{dz^p}l_{n+p}\sum_{r=0}^\infty\frac{a_\gamma^r}{r!}J^r,
\end{multline}
so we se that these sections transform in a way similar to the transformations of the $\theta$-functions
\be\label{eq_sec}
s(z+\gamma)=e^{a_\gamma z+ b_\gamma}s(z)\exp(a_\gamma J),
\ee
and it is clear how to construct rank $M$ vector bundles with this idea. To do that we set $l_n=0$ for $m\ge M$, so that $s(z)$ is an $M$-component row-vector, then $J$ becomes an $M\times M$ matrix with $1$s on the first diagonal above the main diagonal, and zeros otherwise. These sections satisfying $(\ref{eq_sec})$ are holomorphic sections of a rank $M$ indecomposable vector bundle on the torus. If we start with degree $d'$ theta functions, we get a vector bundle of degree $d=Md'$. The space of sections of this bundle is generated by the above sections  (with $0\le n\le M-1$). Its dimension is equal to $d=Md'$.

Let us consider for simplicity the case of $M=2$. We pick a basis of the $\theta$-functions with the type $(a_\gamma,b_\gamma)$ and degree $d'$, and denote them as $\theta_0,\theta_1,\ldots\theta_{d'-1}$. Denoting by $\mathcal{V}$ the corresponding rank 2 bundle, we have a basis of sections of $\mathcal{V}$ denoted by $\sigma_0,\sigma_1,\ldots\sigma_{d-1}$ where $d=2d'$, given by:
\begin{align}
\sigma_j(z)=(\theta_j(z),\theta_j'(z)),\\
\sigma_{d'+j}(z)=(0,\theta_j(z)),
\end{align}
with $0\le j\le d'-1$, where $\theta'(z)=d\theta(z)/dz$.

The texture is constructed by choosing $N$ sections $s^{(1)}(z),s^{(2)}(z),\ldots s^{(N)}(z)$. Because each one can be expanded in the above basis we have
an $N\times d$ matrix $A$, such that
\be
s^{(i)}(z)=\sum_{j=1}^d A_{ij}\sigma_{j}(z).
\ee
We remark that there is a constraint here, namely that $N$ row vectors $s^{(i)}(z)$ with $(1\le i\le N)$ should generate the 2-dimensional row-space at each point $z$. It is satisfied for ``generic'' $A$ matrices. Applying a Pl\"ucker map to this texture we get an $\mathbb{C}P(\tilde{N}-1)$ texture over the torus with $\tilde N=N(N-1)/2$. 

For any pair of sections $(s^{(1)},s^{(2)})$ of $\mathcal{V}$ we denote $s^{(1)}\wedge s^{(2)}$ the section of $\mathrm{Det}\mathcal{V}$ obtained by taking at each $z$ the $2\times 2$ determinant
\begin{equation}
\begin{vmatrix}
s_{11}(z) & s_{12}(z) \\
s_{21}(z) & s_{22}(z).
\end{vmatrix}
\end{equation}
Since $s^{(1)}(z)$ and $s^{(2)}(z)$ both satisfy the equation $(\ref{eq_sec})$ we get
\begin{equation}
s^{(1)}(z+\gamma)\wedge s^{(2)}(z+\gamma)=e^{2(a_\gamma z+b_\gamma)}\mathrm{Det} (e^{a_\gamma J})s^{(1)}(z)\wedge s^{(2)}(z),
\end{equation}
and noticing that $\mathrm{Det}(e^{a_\gamma J})=1$ we obtain
\begin{equation}
s^{(1)}(z+\gamma)\wedge s^{(2)}(z+\gamma)=e^{2(a_\gamma z+b_\gamma)}s^{(1)}(z)\wedge s^{(2)}(z),
\end{equation}
From this transformation properties under $\gamma$ one can see that the sections of $\mathrm{Det}\mathcal{V}$ are $\theta$ functions of charge $d=2d'$
whose type is given by $(\tilde a_\gamma,\tilde b_\gamma)=2(a_\gamma, b_\gamma)$.

Let us introduce an \textit{orthonormal basis} (for the standard hermitean scalar product, think of a lowest Landau level with $2d'=d$ flux quanta): $\tilde\theta_0(z),\tilde\theta_1(z),\ldots\tilde\theta_{d-1}(z)$ for the sections of $\mathrm{Det}\mathcal{V}$. The Pl\"ucker map from $\mathrm{Gr(2,N)}$ to $\mathbb{C}P(\tilde N-1)$ is encoded by the coefficients $B_{j_1j_2;j}$ defined by
\be
\sigma_{j_1}(z)\wedge\sigma_{j_2}(z)=\sum_{j=0}^{d-1}B_{j_1j_2;j}\tilde\theta_j(z),\ \ \ 0\le j_1<j_2\le d-1.
\ee
Given the $N\times d$ matrix $A$ defining the Grassmannian $\mathrm{Gr(2,N)}$ texture, its image under Pl\"ucker embedding is
defined by the $\tilde N\times d$ matrix $\tilde A$:
\begin{equation}
\tilde A_{i_1i_2;j}=\sum_{0\le j_1<j_2\le d-1}
\begin{vmatrix}
A_{i_1j_1} & A_{i_1j_2} \\
A_{i_2j_1} & A_{i_2j_2}
\end{vmatrix}
B_{j_1j_2;j}.
\end{equation}
We recall that $\tilde A$ has the usual meaning for a projective texture
\be
s^{(i_1)}(z)\wedge s^{(i_2)}(z)=\sum_{j=0}^{d-1} \tilde A_{i_1i_2;j}\tilde\theta_j(z), \ \ \ 1\le i< i_2\le N.
\ee
Now, from the optimality condition for the $\mathbb{C}P(N-1)$ models, if the basis $\{\tilde\theta_j(z)\}_{0\le j_1\le d-1}$ is orthonormal the optimality condition in terms of the matrix $\tilde A$ is given by the equation
\be
\tilde A^{\dagger}\tilde A=1,
\ee
which is one of the central results of this paper. In terms of matrix $A$ this optimality equation can be written as
\be
\tilde A^{\dagger}\tilde A=B^{\dagger}M(A)B=1,
\ee
where the matrix $M(A)$ is defined as
\be
M(A)_{j_1j_2;j'_1j'_2}=
\begin{vmatrix}
(A^{\dagger}A)_{j_1j'_1} & (A^{\dagger}A)_{j_1j'_2} \\
(A^{\dagger}A)_{j_2j'_1}& (A^{\dagger}A)_{j_2j'_2}
\end{vmatrix}.
\ee
So the optimality condition involves only hermitian scalar products of the $N$-dimensional columns of $A$. These scalar products are clearly
gauge-invariant, and we can regard these scalar products as our basic degrees of freedom.

\section{Taking an $M$-sheeted covering of the $(\gamma_1,\gamma_2)$ torus. Case II.}

The idea is to start from a line bundle ${L}^{(M)}$ of degree $d$ over the torus ${T}^{(M)}=\mathbb{C}/\mathbb{Z}\gamma_1\oplus M\mathbb{Z}\gamma_2$
\begin{figure}[t]
\epsfig{file=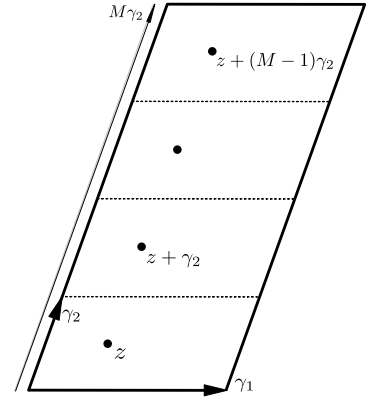,width=5cm}
\caption{An $M$-sheeted covering of a torus with periods $(\gamma_1,\gamma_2)$}
\end{figure}
This torus $T^{(M)}$ is an $M$-sheeted covering of the basic torus $T=\mathbb{C}/\mathbb{Z}\gamma_1\oplus\mathbb{Z}\gamma_2$. We denote by $\pi_M$ the projejction from $T^{(M)}$ to $T$. For $z\in T$ (i.e. $z$ belongs to fundamental parallelogram), the inverse image $\pi^{-1}_M(z)=\{z, z+\gamma_2,\ldots,z+(M-1)\gamma_2\}$. Given $L^{(M)}$ over $T^{(M)}$, we consider the push-forward bundle $\pi_{M*}L^{(M)}$over $T$ denoted by $\mathcal{V}$. It is a rank $M$, degree $d$ vector bundle over $T$, and Atiyah has shown that it is indecomposable (but only when $M$ and $d$ are mutually prime numbers). Its space of holomorphic sections $H^{0}(\mathcal{V},T)$ is the same as $H^{0}(L^{(M)},T^{(M)})$. The latter is the space of $\theta$-functions of degree $d$ over the $T^{(M)}$ torus. For any such $\theta$-function its push-forward $\pi_{M*}\theta$ can be seen as length $M$ row-vector
\be\label{eq10}
\pi_{M*}\theta(z)=(\theta(z),\theta(z+\gamma_2),\ldots,\theta(z+(M-1)\gamma_2)).
\ee
Here, the type of these $\theta$-functions is chosen such that $a_{\gamma_1}=0$. Then we have the following relations
\begin{align}\label{eq11}
&\theta(z+\gamma_1)=e^{b_{\gamma_1}}\theta(z),\\
&\theta(z+M\gamma_2)=e^{a_{M\gamma_2}z+b_{M\gamma_2}}\theta(z).
\end{align}

From the equations $(\ref{eq10})$ and $(\ref{eq11})$ we get transition functions ("factor of automorphy") of the $\mathcal{V}$ bundle over $T$"
\begin{align}
&\pi_{M*}\theta(z+\gamma_1)=e^{b_{\gamma_1}}\theta(z) 
&\pi_{M*}\theta(z+\gamma_2)=\pi_{M*}\theta(z)A(z),
\end{align}
with matrix $A(z)$ defined as
\begin{equation}\label{eq12}
A(z)=
\begin{pmatrix}
0 & \ldots & 0 & e^{a_{M\gamma_2 z}+b_{M\gamma_2}}\\
1 & \ddots &  & 0\\
0 & 1 & \ldots & \vdots \\
0 & 0 & \ldots & 0
\end{pmatrix}.
\end{equation}
Let us chose the standard basis $\theta^{(M)}_j(z)$, $(0\leqslant j\leqslant d-1)$ of sections of $L^{(M)}$ over $T^{(M)}$. We get a basis
$\sigma_j(z)$, $(0\leqslant j\leqslant d-1)$ of $H^{0}(\mathcal{V},T)$ simply by taking $\sigma_j\equiv\pi_{M*}\theta^{(M)}_j$.

The Pl\"ucker map involves taking antisymmetrized product of the form $\sigma_{j_1}(z)\wedge\sigma_{j_2}(z)\wedge\ldots\wedge\sigma_{j_M}(z)$, $(0\leqslant j_1<\ldots< j_M\leqslant d-1)$. This may be expressed as $M\times M$ determinant:
\begin{equation*}
\sigma_{j_1}\wedge\ldots\wedge\sigma_{j_M}=
\begin{vmatrix}
\theta_{j_1}^{(M)}(z) & \theta_{j_1}^{(M)}(z+\gamma_2) & \ldots & \theta_{j_1}^{(M)}(z+(M-1)\gamma_2)\\[4pt]
\theta_{j_2}^{(M)}(z) & \theta_{j_2}^{(M)}(z+\gamma_2) & \ldots & \theta_{j_2}^{(M)}(z+(M-1)\gamma_2)\\
\vdots& \vdots & \vdots & \vdots \\
\theta_{j_M}^{(M)}(z) & \theta_{j_M}^{(M)}(z+\gamma_2) & \ldots & \theta_{j_M}^{(M)}(z+(M-1)\gamma_2)
\end{vmatrix}.
\end{equation*}
It is clear that these behave as degree $d$ theta functions over the $T$ torus:
\begin{align}\label{eq14}
&\sigma_{j_1}\wedge\ldots\wedge\sigma_{j_M}(z+\gamma_1)=e^{M b_{\gamma_1}}\sigma_{j_1}\wedge\ldots\wedge\sigma_{j_M}(z)\\
&\sigma_{j_1}\wedge\ldots\wedge\sigma_{j_M}(z+\gamma_2)=(-1)^{M-1}e^{a_{M\gamma_2} z+b_{M\gamma_2}}\sigma_{j_1}\wedge\ldots\wedge\sigma_{j_M}(z).\nonumber
\end{align}
Introducing an \textit{orthonormal basis} of $\theta$ functions of the corresponding type over the $T$ torus denoted by $\tilde\theta_j(z)$, $(0\leqslant j\leqslant d-1)$, the Pl\"ucker map is expressed through the decomposition:
\be
\sigma_{j_1}(z)\wedge\sigma_{j_2}(z)\wedge\ldots\wedge\sigma_{j_M}(z)=\sum_{j=0}^{d-1}B_{j_1\ldots j_M;j}\ \tilde\theta_j(z).
\ee
The rest of the discussion the same as in the case above.

\section{Remarks on general indecomposable vector bundles}
In the general case (see e.g. Th. 5.21 in the paper of O.~Iena, arXiv:1009.3230) one can write $\mathrm{gcd}(M,d)=h$, and we can define $M=h M'$ and $d=h d'$ with $\mathrm{gcd}(M',d')=1$. The Case I above corresponds to $h=M$, so that $M'=1$ and $d=Md'$, and specifically for $M=2$ it corresponds to even topological charge $d$. The Case II is applicable when $h=1$ so that $M'=M$ and $d'=d$. For $M=2$ this corresponds to odd charge $d$. So for $M=2$ we have to treat the cases of even and odd charge separately according to discussion presented for Case I, and Case II respectively. 

For $M$ a \textit{prime} number, the same dichotomy holds. When $M$ is not prime, and $h\ne 1$ and $h\ne M$ one has to use a 2-step construction. We start with a line bundle $L^{(M')}$ of degree $d'$ over the $T^{(M')}$ torus. Using Case I we construct a vector bundle $\mathcal{V}^{(M')}$ of rank $h$ and degree $d$ over this torus. In the second step we apply the push-forward construction (generalization of Case II) where the line bundle $L^{(M')}$ is replaced by the rank $h$ bundle $\mathcal{V}^{(M')}$) to get a vector bundle $\mathcal{V}=\pi_{M'*}\mathcal{V}^{(M')}$ of rank $M'h=M$ and degree $d$ over $T$.

\section{Explicit values of $B$ coefficients for $M=2$}

\subsection{Case of even $d=2d'$}
We are in the setting of Case I with M=2. We need to define $\theta$ functions of degree $d'$ over the $(\gamma_1,\gamma_2)$ torus. Without loss of generality we can choose $\gamma_1=\pi$ and $\gamma_2=\pi\tau$ with $\mathrm{Im}\ \tau>0$. As usual, it is convenient to introduce $q=\exp(i\pi\tau)$, so $|q|<1$. We choose the following type of the $\theta$ functions defined by $a_{\gamma_1}=0,\ a_{\gamma_2}=-2 i d'$ and $b_{\gamma_1}=0,\ b_{\gamma_2}=0$. This gives the basis of $\theta$ functions:
\be
\theta_p(z)=\sum_{n\in\mathbb{Z}}q^{d'(n-p/d')(n-p/d'-1)}e^{2i(n d'-p)z}.
\ee
Recall that the elementary translations preserving this type are given by:
\begin{align}
&\mathcal{T}_{\gamma_1/d'}\theta(z)=\theta(z-\gamma_1/d')\\
&\mathcal{T}_{\gamma_2/d'}\theta(z)=q^{1+1/d'}e^{-2 i z}\ \theta(z-\gamma_2/d'),
\end{align}
then
\begin{align}
&\mathcal{T}_{\gamma_1/d'}\theta_p=e^{i 2\pi p/d'}\theta_p\\
&\mathcal{T}_{\gamma_2/d'}\theta_p=\theta_{p+1}.
\end{align}

We construct the rank $2$ bundle $\mathcal{V}$ as explained in Case I by taking these $\theta$-functions and their first derivatives. The line bundle
$\mathrm{Det}\mathcal{V}$ defines $\theta$-functions of degree $d=2d'$ over the $(\gamma_1,\gamma_2)$ torus. Their type is defined by $\tilde a_{\gamma_1}=0,\ \tilde a_{\gamma_2}=-2 i d$ and $\tilde b_{\gamma_1}=0,\ \tilde b_{\gamma_2}=0$. Note that $\theta_{p+d'}=\theta_p$, so $(\mathcal{T}_{\gamma_2/d'})^{d'}=\mathbb{I}$ (span of $\theta$ functions). These $\theta$-functions have the following basis
\begin{equation}
\tilde\theta_p(z)=\sum_{n\in\mathbb{Z}}q^{d(n-p/d)(n-p/d-1)}e^{2 i (n d -p)z},
\end{equation}
note that $\tilde\theta_{p+d}=\tilde\theta_p$.

We wish to compute the expansion coefficients $B_{j_1 j_2;j}$. For this we need to decompose products of $\theta_{j_1}\theta_{j_2}$ and the Wronskians of $\theta_{j_1}$ and $\theta_{j_2}$ on the $\tilde\theta_{j}$ basis. We get

\begin{multline}
\theta_{j_1}(z)\theta_{j_2}(z)=A(j_2-j_1)\tilde\theta_{j_1+j_2}(z)\\+A(j_2-j_1+d')\tilde\theta_{j_1+j_2-d'}(z)
\end{multline}
with 
\begin{equation}
A(j)=\sum_{m\in\mathbb{Z}}q^{d(m+j/d)^2},
\end{equation} 
and we note that $A(-j)=A(j)$.

We note that in the large $d$ limit the amplitude $A(j)$ is $d$-periodic and peaks at $j\in d\ \mathbb{Z}$. The width of these peaks is of the order of $\sqrt{d}$ which is much smaller than the period $d$ at large $d$. So, the collection of $B_{j_1 j_2;j}$ coefficients becomes rather sparse in this limit.

We also need expressions for the Wronskian of the theta-functions, which is given by
\begin{multline}
\theta_{j_1}\frac{d\theta_{j_2}}{dz}-\frac{d\theta_{j_1}}{dz}\theta_{j_2}=-2 i d [B(j_2-j_1)\tilde{\theta}_{j_1+j_2}(z)\\+B(j_2-j_1+d')\tilde\theta_{j_1+j_2-d'}(z)],
\end{multline}
with
\be
B(j)=\sum_{m\in\mathbb{Z}}(m+j/d)q^{d(m+j/d)^2}.
\ee
In the large $d$ limit $B(j)$ has a ``dipolar'' profile (as for the derivative of the Gaussian), with the distance of the order of $\sqrt d$ between the local minimum
and the local maximum of $B(j)$, located symmetrically with respect to $j=0$. Note that $B(-j)=-B(j)$.

\subsection{Explicit expressions for the coefficients of the expansion for $M=2$ in the case of odd $d$}  
We use the 2-sheeted covering $T^{(2)}=\mathbb{C}/\mathbb{Z}\gamma_1+2\mathbb{Z}\gamma_2$ of the $(\gamma_1,\gamma_2)$ torus.
On the torus $T^{(M)}$ with $(M=2)$ we introduce $\theta$ functions of degree $d$, where type is defined by $a_{\gamma_1}=0,\ a_{M\gamma_2}=-2 i d$ and $b_{\gamma_1}=0,\ b_{\gamma_2}=0$. A basis for these functions is given by
\be
\theta^{(M)}_j(z)=\sum_{n\in\mathbb Z}q^{M d(n-j/d)(n-j/d-1)}e^{2 i (n d-j)z}.
\ee
On the $(\gamma_1,\gamma_2)$ torus we use almost the same $\tilde\theta_j(z)$ basis of degree $d$ theta functions as before, but with a slightly modified type, so $\tilde b_{\gamma_2}=i\pi$, and we get
\be
\tilde\theta_j(z)=\sum_{n\in\mathbb{Z}}(-1)^{n}q^{d(n-j/d)(n-j/d-1)}e^{2 i (n d-j)z}.
\ee
We have seen that
\be
\theta_{j_1}(z)\wedge\theta_{j_2}(z)=
\begin{vmatrix}
\theta^{(2)}_{j_1}(z) & \theta^{(2)}_{j_1}(z+\gamma_2)\\[4pt]
\theta^{(2)}_{j_2}(z) & \theta^{(2)}_{j_2}(z+\gamma_2),
\end{vmatrix}
\ee
and we get
\be\label{eq27}
\theta_{j_1}(z)\wedge\theta_{j_2}(z)=C(j_2-j_1)\tilde\theta_{j_1+j_2}(z),
\ee
where
\be
C(j)=\sum_{m\in\mathbb{Z}}(-1)^{m}q^{d(m+j/d)(m+j/d-1)}.
\ee

\section{Automorphisms of the indecomposable bundles on the torus}

\subsection{Construction using derivatives of the $\theta$-functions}
We start from the automorphy factors given by the equation
\be\label{automorphism}
s(z+\gamma)=e^{a_\gamma z+b_\gamma}\exp(a_\gamma J),
\ee
with $J$ being the following matrix
\be
J=
\begin{pmatrix}
0 & 1 & \ldots & 0\\
0 & 0 & 1 & \ldots \\
\vdots & \vdots & \vdots & \vdots \\
0 & \ldots & 0 &1\\
0 & \ldots & \ldots & 0
\end{pmatrix}.
\ee
An automorphism of such bundle is defined by an invertible $M\times M$ matrix $\Phi(z)$, holomorphic in $z$. Applying $\Phi(z)$ to the right of the row vector $s(z)$ gives $t(z)=s(z)\Phi(z)$, which should also satisfy $(\ref{automorphism})$. Therefore, we wish to impose the condition:
\be\label{eq_phi}
\exp(a_\gamma J)\Phi(z+\gamma)=\Phi(z)\exp(a_\gamma J).
\ee
Since $a_{\gamma_1}=0$, we have $\Phi(z+\gamma_1)=\Phi(z)$. Chosing $\gamma_1=\pi$, we set $u=e^{2 i z}$. Any translation $z\to z+n \gamma_1,\ n\in\mathbb{Z}$ leaves $u$ invariant. Therefore, we may look for $\Phi$ as a holomorphic function of $u$. Note that $u\in\mathbb{C}\backslash \{0\}$, so we may write $\Phi$ as: 
\be
\Phi(u)=\sum_{k\in\mathbb{Z}}u^k \Phi_k,
\ee
where $\Phi_k$ is an $M\times M$ matrix. We will be using previous notations $a_{\gamma_2}=-2 i d'$, and $\gamma_2=\pi\tau$, and $q=\exp{i\pi\tau}$, note that O.~Iena arXiv:1009.3230 uses a different definition for $q_{Iena}=\exp(2 i\pi\tau)$, then we see that if $z\to z+\gamma_2$ then $u\to q^2 u$.  Then for $\gamma=\gamma_2$ the equation $(\ref{eq_phi})$ gives
\be
\Phi(q^2 u)=e^{2 i d' J} \Phi(u) e^{-2 i d' J},
\ee
which translates into equation
\be\label{phik}
q^{2k}\Phi_k=e^{2 i d' J}\Phi_k e^{-2 i d' J}.
\ee
Note that since $J^M=0$ we have
\be
e^{2 i d' J}=I + \sum_{p=1}^{M-1}\frac{(2 i d')^p}{p!} J^p \equiv I+\hat N,
\ee
where the $M\times M$ matrix automorphism $\hat N$ also satisfies 
the condition $\hat N^M=0$.

\textit{Lemma:} consider the matrix equation for $\Phi$: $c\,\Phi(1+\hat N)=(1+\hat N)\Phi, \ c\in\mathbb{Z}$. If $c\ne 1$ the only solution is $\Phi=0$. If $c=1$, then $\Phi$ is a linear combination of $I, J, J^2,\ldots, J^{M-1}$.

Let us check this statement. We set $\alpha\equiv 2 i d'$ and consider 
\be
\hat N^p=(\alpha J+\frac{\alpha^2}{2}J^2+\ldots+\frac{\alpha^{M-1}}{(M-1)!}J^{M-1})^p,
\ee
so that
\be
\hat N^p=\alpha^p J^p+ (\ldots)J^{p+1}+\ldots+(\ldots)J^{M-1},\ \ p\le M-1.
\ee
This shows that $\hat N^{M-1}\ne 0$ but $\hat N^{M}=0$. So there exists a basis $e_1,\ldots,e_M$ of $\mathbb{C}^{M}$ in which $\hat N$ has the canonical Jordan form, i.e.
\be
\hat N e_1=0,\ \hat N e_2=e_1,\ldots, \hat N e_M=e_{M-1}.
\ee

Let us first consider the case $c\ne 1$, and we set $\Phi(e_1)=x_1 e_1+\ldots+x_M e_M$, then $c\Phi(e_1)=(I+\hat N)\Phi(e_1)$,
so
\begin{align*}
&c x_1= x_1+x_2,\\
&c x_2=x_2+x_3,\\
&\vdots\\
&c x_{M-1}=x_{M-1}+x_M,\\
&c x_M=x_M.
\end{align*}
If $c\ne 1$ the last equation gives $x_M=0$, but then $x_{M-1}=0,\ldots x_2=0$, and $x_1=0$ so
$\Phi(e_1)=0$. Since $\hat N e_2=e_1$ and using $\Phi(e_1)=0$ we get $c \Phi(e_2)=(1+\hat N)\Phi(e_2)$ so also $\Phi(e_2)=0$. Then the same reasoning shows that $\Phi(e_3),\ldots,\Phi(e_M)=0$. Therefore, $c\ne 1$ implies that $\Phi=0$ is the only solution.

Now consider the case of $c=1$. We wish to determine the matrices which commute with $\hat N$. We shall show that such matrices also commute with $J$. For this we may express $J$ as a polynomial in the powers of $\hat N$ of degree $M$. For all $\alpha \in \mathbb{R}$ we have
\be
\sum_{p=1}^{M-1}\frac{(-1)^{p-1}}{p}(e^{\alpha J}-I)^p=\alpha J.
\label{J_versus_hatN}
\ee
We take the derivative of the L.H.S.~ with respect to $\alpha$. This gives:
\be
J\sum_{p=0}^{M-2}e^{\alpha J}(-1)^{p}(e^{\alpha J}-I)^p=J e^{\alpha J}\sum_{p=0}^{M-1}(-1)^p (e^{\alpha J}-1)^p,
\ee
which is due to $(e^{\alpha J}-1)^M=0$, so taking derivative with respect to $\alpha$ gives $J e^{\alpha J}(e^{\alpha J}-1)^{M-1}=0$. Then
\be\label{eq34}
\sum_{p=0}^{M-1}(-1)^p(e^{\alpha J}-1)^p=\sum_{p=0}^{M-1}(-1)^p\hat N^p,
\ee
So we see that $(1+\hat N)\sum_{p=0}^{M-1}(-1)^p\hat N^p=I+(-1)^{M-1}\hat N^M=I$, and because $e^{\alpha J}=1+\hat N$,
both sides of equation $(\ref{J_versus_hatN})$ have the same $\alpha$-derivative. Because they also coincide for $\alpha=0$, this proves $(\ref{J_versus_hatN})$. 

We note that $(\ref{J_versus_hatN})$ is of course the series expansion of $\log[I+(e^{\alpha J}-1)]$. It is often proved only for diagonalisable matrices, but since $e^{\alpha J}=1+\hat N$, with $\hat N$ nilpotent, $e^{\alpha J}$ is not diagonalisable. This is why we showed a direct check of this equation in our case. Now, from  $(\ref{J_versus_hatN})$ we see that if $[\Phi,\hat N]=0$, then $[\Phi,\hat N^p]=0$ for all $p\in\mathbb{N}$, and then $[\Phi,J]=0$. Let us denote by $(b_1,b_2,\ldots,b_m)$ the canonical basis of $\mathbb{C}^M$. Then $J b_p=b_{p-1}$ for $p\ge 2$ and $J b_1=0$. It is easy to show, by increasing recursion on $p$, that there are $c$-numbers $\lambda_0,\lambda_1,\ldots,\lambda_{M-1}$ such that:
\be
\Phi b_p=\lambda_0 b_p+\lambda_1 b_{p-1} +\lambda_2 b_{p-2}+\ldots+\lambda_{p-1}b_1.
\ee
Then 
\be
\Phi=\lambda_0 I+\lambda_1 J+\lambda_2 J^2+\ldots + \lambda_{M-1}J^{M-1}.
\label{Phi_versus_J}
\ee

Let us now return to equation $(\ref{phik})$, since $\mathrm{Im}\tau>0$ we have $|q|<1$ so $q^{2k}\ne 1$ if $k\ne 0$. The Lemma shows that $\Phi_k=0$ for $k\ne 0$. For $k=0$ it shows that $\Phi_0$ is a linear combination of the form $(\ref{Phi_versus_J})$. We have therefore shown that:

The only automorphisms of the vector bundle constructed by the procedure of Case I (involving derivatives of the theta functions) are obtained by multiplying $s(z)$ on the right by a constant $M\times M$ matrix $\Phi_0$ as in $(\ref{Phi_versus_J})$, explicitely
\be
\begin{bmatrix}
\lambda_0 & \lambda_1 & \lambda_2 & \ldots & \lambda_{M-1}\\
0 & \lambda_0 & \lambda_1 & \ldots & \vdots\\
0 & 0 & \lambda_0 & \ldots & \vdots\\
\vdots & \vdots & \vdots & \vdots & \vdots\\
0 & 0 & 0 & 0 & \lambda_0
\end{bmatrix},
\ee
with $\lambda_0,\ldots,\lambda_{M-1}$ complex constants.

\subsection{Pushing forward line bundles on an $M$-sheeted covering of the torus}
Let us keep notation of section Case II, and choose $b_{\gamma_1}=0$ to simplify the
discussion. Sections of $\pi_{M*}L^{(M)}$ are then invariant under $\gamma_1$ translations.
As above, automorphisms are represented by $\Phi(z)$, holomorphic and invertible $M\times M$ matrix, such that $\Phi(z+\gamma_1)=\Phi(z)$. So we may write $\Phi$ as a Laurent series in $u=\exp(2 i z)$. We have now to express compatibility of $\Phi(z)$ with $\gamma_2$ translations. Using the $M\times M$ matrix $A(z)$ introduced in equation $(\ref{eq12})$, compatibility of $\Phi$ requires
\be\label{eq35}
A(z)\Phi(z+\gamma_2)=\Phi(z)A(z).
\ee
where 
\begin{equation}
A(z)=
\begin{pmatrix}
0 & \ldots & 0 & e^{a_{M\gamma_2}z+b_{M\gamma_2}}\\
1 & \ddots &  & 0\\
0 & 1 & \ldots & \vdots \\
0 & 0 & \ldots & 0
\end{pmatrix}.
\end{equation}
This is a cyclic matrix, so taking $M$-fold product of such matrices gives a diagonal matrix. In particlular
\begin{multline}
D(z)\equiv A(z)A(z+\gamma_2)\ldots A(z+(M-1)\gamma_2)=\\
\begin{pmatrix}
\lambda(z) & 0 & \ldots & 0\\
0 & \lambda(z+\gamma_2) & \ldots & \vdots\\
\vdots & \vdots & \vdots & \vdots\\
0 & \ldots & \ldots & \lambda(z+(M-1)\gamma_2)
\end{pmatrix}.
\end{multline}
Repeated application of equation $(\ref{eq35})$ gives
\begin{align*}
\Phi(z)D(z) &= \Phi(z) A(z)A(z+\gamma_2)\ldots A(z+(M-1)\gamma_2)\\
& =  A(z) \Phi(z+\gamma_2)A(z+\gamma_2)\ldots A(z+(M-1)\gamma_2)\\
& \vdots\\
\Phi(z)D(z) &=D(z)\Phi(z+M\gamma_2),
\end{align*}
so
\be
\Phi_{jk}(z)\ \lambda(z+(k-1)\gamma_2)=\lambda(z+(j-1)\gamma_2)\Phi_{jk}(z+M\gamma_2),
\ee
and we have
\be\label{eq38}
\Phi_{jk}(z+M\gamma_2)=\exp(a_{M\gamma_2}\gamma_2 (k-j))\Phi_{jk}(z).
\ee
We recall that $a_{M\gamma_2}=-2 i d$ and $\gamma_2=\pi\tau$. With $u=e^{2 i z}$ and $q=e^{i\pi\tau}$ equation $(\ref{eq38})$ reads:
\be
\Phi_{jk}(q^{2M}u)=q^{2d(j-k)}\Phi_{jk}(u).
\ee
Writing $\Phi_{jk}$ in a Laurent series $\Phi_{jk}=\sum_{n\in\mathbb{Z}}\Phi_{jk}^{(n)}u^n$, we have
\be\label{eq40}
(q^{2 M n}-q^{2 d (j-k)})\Phi_{jk}^{(n)}=0.
\ee
So $\Phi_{jk}(u)\ne 0$ only if $d(j-k)\in M\,\mathbb{Z}$. When $d$ and $M$ are relatively prime, as for example when $M=2$ and $d$ is odd, this happens only for $j=k$, i.e.~for diagonal elements of $\Phi(u)$. When $j=k$, only $n=0$ makes the L.H.S.~ of $(\ref{eq40})$ vanish, since $0<|q|<1$. So when $M$ and $d$ are mutually prime, the only automorphisms of $\pi_{M*}L^{(M)}$ are trivial, i.e.~$\Phi(z)=const\times I$.

\section{$B^{\dagger}B\propto I$ for an $M$-sheeted covering of the $(\gamma_1,\gamma_2)$ torus}
\subsection{Translations of line bundles $L^{(M)}$ over $(\gamma_1,M\gamma_2)$ and $\tilde L$ over $(\gamma_1,\gamma_2)$}
Suppose that sections satisfy $s(z+\gamma)=e^{a_\gamma z+b_\gamma}s(z)$. Here $\gamma=M\gamma_2$ or $\gamma=\gamma_2$. Type preserving translations have the form
\be
[\mathcal{T}_{\gamma/d} s](z)=\lambda e^{\mu z}s(z-\gamma/d),
\ee
with the constraint $\exp(\mu\gamma-a_{\gamma}\gamma/d)=1$, and we take $\mu=a_\gamma/d$. In order to find $\lambda$ we impose that $(\mathcal{T}_{\gamma/d})^d=I$, so
\begin{multline*}
[\mathcal{T}_{\gamma/d}s](z+\gamma)=\lambda^d\exp(\frac{a_\gamma}{d}(z+\gamma+z+\frac{d-1}{d}\gamma+\ldots+z+\frac{\gamma}{d}))s(z)\\
=\lambda^d\exp[\frac{a_\gamma}{d}(d z+\frac{d+1}{2}\gamma)]s(z).
\end{multline*}
Because $[\mathcal{T}_{\gamma/d}]^d s= s$ the L.H.S.~is equal to $e^{a_\gamma z+b_\gamma}s(z)$ so we have 
\be
\lambda=\exp(\frac{b_\gamma}{d}-a_{\gamma}\gamma\frac{(d+1)}{2d^2}).
\ee
For the line bundle $L^{(M)}$ over the $(\gamma_1,M\gamma_2)$ torus, $\gamma=M\gamma_2$ and $b_\gamma=0$ then
\be
\lambda=\exp(-\frac{(d+1)M}{2d^2}a_{M\gamma_2}\gamma_2).
\ee
For the line bundle $\tilde L=\mathcal{D}et(\pi_{*}L^{(M)})$ over the $(\gamma_1,\gamma_2)$ torus, $\gamma=\gamma_2$. Equation $(\ref{eq14})$ shows that the multiplicative factor $a_\gamma=a_{M\gamma_2}$ and $\exp(b_{\gamma_2})=(-1)^{M-1}$, then
\be
\tilde{\lambda}=(-1)^{(M-1)/d}\exp(-\frac{d+1}{2d^2}a_{M\gamma_2}\gamma_2).
\ee
On $L^{(M)}$ we have a basis of theta functions $\theta_j^{(M)}(z)$ with $j\in\{0,1,\ldots,d-1\}$ with the properties $\mathcal{T}^{(M)}_{M\gamma_2/d}\theta_j^{(M)}=\theta_{j+1}^{(M)}$ and $\theta_{j+d}^{(M)}=\theta_j^{(M)}$. These theta functions are characterised by the effect of translations by $\gamma_1/d$ in the following way
\be
\mathcal{T}^{(M)}_{\gamma_1/d}\theta_j^{(M)}(z)=\theta_j^{(M)}(z-\gamma_1/d)=e^{i\frac{2\pi}{d}j}\theta^{(M)}_j(z).
\ee
One can show that
\begin{multline}
(\tilde{\mathcal{T}}_{\gamma_1/d}\sigma_{j_1}\wedge\ldots\wedge\sigma_{j_M})(z)=\sigma_{j_1}\wedge\ldots\wedge\sigma_{j_M}(z-\gamma_1/d)\\
=e^{i\frac{2\pi}{d}(j_1+j_2+\ldots+j_M)}\sigma_{j_1}\wedge\ldots\wedge\sigma_{j_M}(z).
\end{multline}
Since $\tilde{\mathcal{T}}_{\gamma_1/d}\tilde\theta_j=e^{i\frac{2\pi}{d}j}\tilde{\theta}_j$, we have
\be\label{eq44}
\sigma_{j_1}\wedge\ldots\wedge\sigma_{j_M}=c(j_1,\ldots,j_M)\tilde\theta_{j_1+j_2+\ldots+j_M},
\ee
where $c(j_1,\ldots,j_M)\in\mathbb{C}$. This is, of course a generalisation of the equation $(\ref{eq27})$ obtained previously for $M=2$. The translational invariance observed in $(\ref{eq27})$ is also valid in the general case:
\be\label{eq45}
|c(j_1+1,j_2+1,\ldots,j_M+1)|=|c(j_1,j_2,\ldots,j_M)|.
\ee
Let us establish this result. Using $\theta^{(M)}_{j+1}(z)=\lambda\exp(a_{M\gamma_2}z/d)\theta^{(M)}_j(z-\frac{M\gamma_2}{d})$, and the explicit  
determinantal expression before $(\ref{eq27})$,
\begin{multline}\label{eq46}
\sigma_{j_1+1}\wedge\ldots\wedge\sigma_{j_M+1}(z)=\lambda^M\exp(\frac{a_{M\gamma_2}}{d}(M z+\frac{M(M-1)}{2}\gamma_2))\\
\times\sigma_{j_1}\wedge\ldots\wedge\sigma_{j_M}(z-M\gamma_2/{d}).
\end{multline}
We wish to compute the R.H.S.~of equation $(\ref{eq46})$ with $([\tilde{\mathcal{T}}_{\gamma_2/d}]^M \sigma_{j_1}\wedge\ldots\wedge\sigma_{j_M})(z)$. Pick any section $\tilde\theta$ of $\tilde L$ over the $(\gamma_1,\gamma_2)$ torus. Then:
\begin{multline}
([\tilde{\mathcal{T}}_{\gamma_2/d}]^M \tilde\theta)(z)=\tilde\lambda^M\exp[
\frac{a_{M\gamma_2}}{d}(z+(z-\frac{\gamma_2}{d})+\ldots\\
+(z-\frac{(M-1)\gamma_2}{d}))]   \tilde{\theta}(z-M\gamma_2/d),
\end{multline}
or
\begin{multline}
([\tilde{\mathcal{T}}_{\gamma_2/d}]^M \tilde\theta)(z)=\tilde\lambda^M\exp[\frac{a_{M\gamma_2}}{d}
\left(M z-\frac{M(M-1)}{2d}\gamma_2\right)]\\ \times   \tilde{\theta}(z-M\gamma_2/d).
\end{multline}
If we define $\alpha$ by $e^{i\alpha}(-1)^{\frac{M(M-1)}{d}}=1$, it is easy to check that
\be
\lambda^{M}\exp(a_{M\gamma_2}\gamma_2\frac{M(M-1)}{2d})=e^{i\alpha}\tilde\lambda^M\exp(-a_{M\gamma_2}\gamma_2\frac{M(M-1)}{2d^2}).
\ee
From $(\ref{eq44})$, this implies
\be
c(j_1+1,\ldots,j_M+1)=e^{i\alpha}c(j_1,\ldots,j_M).
\ee

\subsection{Application to $B^{\dagger}B$}
Equation $(\ref{eq44})$ shows that $B^{\dagger}B$ is diagonal if we use the $\tilde\theta_j$ basis for $\mathcal{D}et(\pi_{M*}L^{(M)})$. More precisely we have:
\be
(B^{\dagger}B)_{jj'}=\delta_{jj'}\sum_{j_1<\ldots j_M}\delta^{(d)}_{j;j_1+\ldots+j_M}|c(j_1,\ldots,j_M)|^2,
\ee
(where $\delta^{(d)}$ is the Kronecker symbol for integers modulo $d$). This has the form $(B^{\dagger}B)_{jj'}=\delta_{jj'}D_j$ and $(\ref{eq45})$ implies that $D_{j+M}=D_j$. Since we assume $M$ and $d$ relatively prime (so $\pi_{M*} L^{(M)}$ is an indecomposable rank $M$ vector bunddle), this equation shows that $D_j$ is independent of $j$, therefore $B^{\dagger}B=D\times I_{d}$.

\section{Number of independent $SU(N)$-invariant deformation modes on the sphere}

\subsection{Line bundles on the sphere}

These bundles, for which a standard notation is $\mathcal{O}(d)$, are completely determined by their topological charge $d$, assumed to be
a positive integer. A key elementary result is that the space of global holomorphic sections of $\mathcal{O}(d)$ on $S^{2}$ is
realized by polynomials in $z$ with maximal degree equal to $d$, so its dimension is equal to $d+1$.
Physically, this space is a realization of the Hilbert space of a quantum spin $S=d/2$. The first occurence of this realization in physics has
probably been given by Dirac in his analytical diagonalization of the Hamiltonian of a charged quantum particule in the field of a magnetic monopole \cite{Dirac_31}.
An explicit reformulation of the corresponding monopole harmonics in terms of sections of line bundles on the sphere can be found in \cite{Wu_Yang_76}.
Later, starting from Haldane's pioneering paper \cite{Haldane_83}, explicit descriptions of the lowest Landau level wave-functions on the sphere have played a crucial role 
in constructing many-particle wave-functions for fractional quantum Hall states on the sphere. For the reader's convenience, let us recall how these 
$\mathcal{O}(d)$ bundles and their space of sections can be constructed.

We start by viewing the sphere as the complex projective line $\mathbb{C}P(1)$. Denoting the line through $(x_0,x_1)$ in $\mathbb{C}^2$ by $(x_0:x_1)$,
$\mathbb{C}P(1)$ is obtained as the union of two open subsets $U_0$ and $U_1$, which are both in one to one correspondence with the set $\mathbb{C}$
of complex numbers. Specifically, $U_j = \{(x_0:x_1) \in  \mathbb{C}P(1), x_j \neq 0\}$, for $j=0,1$. On $U_0$, a local coordinate may be chosen as
$z=x_1/x_0$. Note that $\mathbb{C}P(1)=U_{0} \bigcup \{(0:1)\}$, where $(0:1)$ is usually called the point at infinity since it corresponds to $z = \infty$. 
On $U_1$, the natural local coordinate is $w=x_0/x_1$. On the intersection $U_0 \bigcap U_1$, we have the relation $zw=1$, so the correspondence 
between $z$ and $w$ is holomorphic. 

The $\mathcal{O}(1)$ bundle is defined as the dual of the tautological bundle over $\mathbb{C}P(1)$. A global section of  $\mathcal{O}(1)$ is therefore
a smooth collection of linear forms on complex lines $(x_0:x_1)$, parametrized by $(x_0:x_1) \in  \mathbb{C}P(1)$. An obvious choice is to restrict a
given linear form $\phi$ on $\mathbb{C}^2$ to each line $(x_0:x_1)$. This line is generated by $(1,z)$ or equivalently by $(w,1)$ whenever $z$ or $w$ are finite.
If $\phi(x_0,x_1)=a_0 x_0 + a_1 x_1$, we set $s^{(0)}(z)=\phi(1,z)=a_0 + a_1 z$ on $U_0$. Likewise, we set 
$s^{(1)}(w)=\phi(w,1)=a_0 w+ a_1 $ on $U_1$. On $U_0 \bigcap U_1$, $s^{(0)}(z)$ and $s^{(1)}(w)$ are related by 
$x_0 s^{(0)}(z)=x_1 s^{(1)}(w)$, so that: 
\be
s^{(0)}(z) =  t(1,z) s^{(1)}(w),
\label{trans_1}
\ee
where $t(1,z)=z$ is the transition function of the $\mathcal{O}(1)$ bundle on $U_0 \bigcap U_1$. 
It is easy to check that pairs of holomorphic functions $s^{(0)}(z)$ and $s^{(1)}(w)$ related by Eq. (\ref{trans_1}) arise from a
linear form $\phi$ on $\mathbb{C}^2$ as described above. So global sections of $\mathcal{O}(1)$ on the sphere are in one
to one correspondence with homogeneous polynomials of degree 1 in $x_0$ and $x_1$, or equivalently, of degree 1 polynomials in a
single variable $z$ or $w$.

For positive integer $d$, the $\mathcal{O}(d)$ bundle is defined as the tensor product of $d$ identical copies of the $\mathcal{O}(1)$ bundle.
The corresponding transition functions are $t(d,z)=t(1,z)^d=z^d$. It is easy to check that global sections of $\mathcal{O}(d)$ on the sphere are in one
to one correspondence with homogeneous polynomials $P(x_0,x_1)$ of degree $d$ in $x_0$ and $x_1$, or equivalently, of degree $d$ polynomials in a
single variable $z$ or $w$. These later polynomials are obtained as $s^{(0)}(z)=P(1,z)=P(x_0,x_1)/x_{0}^d$ on $U_0$ and
$s^{(1)}(w)=P(w,1)=P(x_0,x_1)/x_{1}^d$ on $U_d$.  

 \subsection{Automorphisms of vector bundles on the sphere}

Any rank $M$ vector bundle  $\mathcal{V}$ on $S^{2}$ is of the form 
$\mathcal{V}=\mathcal{O}(d_1) \oplus \mathcal{O}(d_2)... \oplus \mathcal{O}(d_M)$,
with $d_1 + d_2 + ... + d_M = d$ is the total topological charge. It is useful to describe
explicitely automorphisms of $\mathcal{V}$. Such description may be found in \cite{Lomadze_96}.
Let us give here an elementary presentation. The transition function defining $\mathcal{V}$ is given by a rank $M$
diagonal square matrix $t(z)_{ij}=\delta_{ij}\:t(d_i,z)$. An automorphism of $\mathcal{V}$
is defined by two rank $M$ square matrices $A^{(0)}(z)$ and $A^{(1)}(w)$, which are holomorphic
functions of $z$ and $w$, and which are invertible for all $z$ and all $w$. These two matrices are subjected
by the compatibility condition on $U_0 \bigcap U_1$:
\be
A^{(0)}(z) \: t(z) = t(z) \: A^{(1)}(w)
\ee
Using the fact that the matrix $t(z)$ is diagonal, this translates into:
\be
A^{(0)}(z)_{ij}\;z^{(d_{j}-d_{i})}  =  A^{(1)}(w)_{ij}
\label{explicit_connection_A}
\ee
The right-hand side of (\ref{explicit_connection_A}) is regular at $w=0$, which implies that 
$A^{(0)}(z)_{ij}=0$, unless $d_i \ge d_j$.
Because the holomorphic function $A^{(0)}(z)_{ij}$ is bounded by a constant times $|z|^{(d_{i}-d_{j})}$ as
$|z|$  becomes large, it is easy to deduce from the Cauchy formula that $A^{(0)}(z)_{ij}$ is a polynomial in
$z$, whose degree is at most equal to $d_i - d_j$. From (\ref{explicit_connection_A}), we also see that the
same property holds for $A^{(1)}(w)_{ij}$. At this stage, it is convenient to assume that
$d_1 \leq d_2 \leq \cdots d_M$. Since some of these degrees can be equal, we introduce the following notation:
\begin{eqnarray*}
d_1 = d_2 = \cdots = d_{i_1} & = & \tilde{d}_1 \\
d_{i_{1}+1} = d_{i_{1}+2} = \cdots = d_{i_{1}+i_{2}} & = & \tilde{d}_2 \\
& . & \\
d_{i_{1}+i_{2}+ \cdots +i_{p-1}+1} = \cdots = d_{i_{1}+i_{2}+ \cdots +i_{p}} & = & \tilde{d}_p
\end{eqnarray*}
with $\tilde{d}_1 < \tilde{d}_2 < \cdots < \tilde{d}_p$, 
$i_{1}+i_{2}+ \cdots +i_{p}=M$ and $ i_1 \tilde{d}_1 + \cdots + i_p \tilde{d}_p = d_1 + \cdots + d_M = d$, where
$d$ is the total topological charge of the map. It is then convenient to view $A^{(0)}(z)$ as a $p \times p$ block
matrix, denoted by $\tilde{A}^{(0)}(z)$, for which the element $\tilde{A}^{(0)}_{lm}(z)$ is an $i_l \times i_m$ matrix. 
The previous discussion shows that $\tilde{A}^{(0)}_{lm}(z)=0$ if $l<m$ and $\tilde{A}^{(0)}_{lm}(z)$ has polynomial
entries of degree at most equal to $\tilde{d}_{l}-\tilde{d}_{m}$ if $l \geq m$. In particular, diagonal blocks $\tilde{A}^{(0)}_{ll}(z)$
are constant invertible rank $i_l$ square matrices, and $\tilde{A}^{(0)}(z)$ is lower triangular.
It is easy to check that if $\tilde{A}^{(0)}(z)$ satisfies the above conditions,
its inverse  $\tilde{B}^{(0)}(z)$ also satisfies them, so the pair $(A^{(0)}(z), A^{(1)}(w))$ defines an automorphism of the vector
bundle  $\mathcal{V}$.

\subsection{$SU(N)$-invariant deformation modes on the sphere}

Assuming that entries of $\tilde{A}^{\dagger}\tilde{A}$
are independent functions of the entries of $A^{\dagger}A$, and that independent automorphisms of $\mathcal{V}$
give rise to independent small deformations of $A^{\dagger}A$, we get the simple estimate of the number $\mathcal{N}_0$ of 
$SU(N)$-invariant deformation modes for optimal textures on the sphere:
\be
\mathcal{N}_0 = \mathcal{N}_{A^{\dagger}A}-\mathcal{N}_{\tilde{A}^{\dagger}\tilde{A}}-\mathcal{N}_{\mathrm{aut}}
\ee
where $\mathcal{N}_{A^{\dagger}A}$ (resp. $\mathcal{N}_{\tilde{A}^{\dagger}\tilde{A}}$) denotes
the number of independent real parameters involved in $A^{\dagger}A$ (resp. $\tilde{A}^{\dagger}\tilde{A}$) and
$\mathcal{N}_{\mathrm{aut}}$ denotes the number of independent real parameters to describe non-trivial automorphisms
of $\mathcal{V}$. The two underlying assumptions behind this simple estimate are hard to prove (or to disprove) by general arguments,
and their validity depends in general on the actual value of $A^{\dagger}A$. For example, if $A^{\dagger}A$ is equal to the unit matrix, it 
remains invariant under the action of any automorphism! Probably these are reasonable assumptions if $A^{\dagger}A$ describing
optimal textures may be considered as {\em generic}, the later notion being only loosely defined. For this reason, it is important to be 
able to produce explicit solutions, at least for small values of $M$, $N$, and $d$, in order to test these assumptions.

To evaluate the number of independent parameters in $A^{\dagger}A$, we first recall that $A$ is an $N \times D$ matrix, where $D$ is
the number of independent global sections of $\mathcal{V}$. Because the sections of the $\mathcal{O}(d)$ line bundle are polynomials
of degree $d$, they depend on $d+1$ complex parameters. Therefore, since $\mathcal{V}$ decomposes as a direct sum of such line bundles,
$D=(d_1 + 1) + (d_2 + 1)+ \cdots + (d_M + 1)=d+M$. Since entries of $A^{\dagger}A$ can be interpreted as hermitian scalar products in $\mathbb{C}^N$
between the $D$ columns of $A$, we have to distinguish according to whether $D \leq N$ or $D > N$. In the first case, the columns of $A$ are
generically linearly independent. It is easy to see that any positive hermitian square matrix of size $D$ and rank $D$ can be realized as an
$A^{\dagger}A$, and that $A^{\dagger}A$ determines $A$ modulo global $SU(N)$ transformations. Therefore, taking into account the fact that
$A^{\dagger}A$ is hermitian, it involves $\mathcal{N}_{A^{\dagger}A}=D^2=(d+M)^2$ real parameters. In the second case $D > N$ implies that 
$A^{\dagger}A$ cannot be of maximal rank $D$, but is rank is limited by $N$, which is smaller than $D$. In the generic case, the rank of
 $A^{\dagger}A$ is equal to $N$. Assuming that the first $N$ columns of $A$ are linearly independent, they are determined, up to a global
 $SU(N)$ transformation,  by the submatrix $(A^{\dagger}A)_{ij}$ for $1 \leq i,j \leq N$, which provides $N^2$ independent real parameters.
 Once its first $N$ columns are known, the remaining $n=D-N$ columns of $A$ are determined by their overlaps with the first $N$ columns,
 which brings $2N(D-N)$ new independent real parameters. Therefore, we see that $A^{\dagger}A$ involves only
 $N^2 + 2N(D-N)=D^2-(D-N)^2=D^2-n^2$ independent real parameters. As in the first case, $A^{\dagger}A$ still determines $A$ modulo global $SU(N)$ transformations.
 
 Regarding $\tilde{A}^{\dagger}\tilde{A}$, it is an hermitian square matrix of linear size $\tilde{d}$, where $\tilde{d}$ denotes the number of
 independent parameters involved in global sections of $\mathrm{Det} \mathcal{V} \simeq \mathcal{O}(d)$. Then $\tilde{d}=d+1$.
 $\tilde{A}^{\dagger}\tilde{A}$  involves $\tilde{d}^2$ independent real parameters,
 so $\mathcal{N}_{\tilde{A}^{\dagger}\tilde{A}}=\tilde{d}^2=(d+1)^2$. At this stage, not taking into account the reduction in the number of physical
 degrees of freedom due to automorphisms, we have the simple formula, in case $d+M \leq N$:
 \be
\mathcal{N}_{A^{\dagger}A}-\mathcal{N}_{\tilde{A}^{\dagger}\tilde{A}} =(M-1)(2d+M+1) 
\label{basic_counting} 
\ee
When $d+M > N$, and $n=D-N$, this number has to be subtracted by $n^2$.

From the explicit description of automorphisms given above, it is easy to see that they depend on 
$\sum_{m \leq l}(\tilde{d}_{l}-\tilde{d}_{l}+1)i_{l}i_{m}$ complex parameters. Removing the trivial automorphism
(proportional to the identity matrix), and multiplying by 2 to count real parameters, we get:
\be
\mathcal{N}_{\mathrm{aut}} = 2 \left(\sum_{m \leq l}(\tilde{d}_{l}-\tilde{d}_{l}+1)i_{l}i_{m}-1\right)
\label{N_aut_sphere} 
\ee
These general expressions simplify greatly when $M=2$. We have two cases, either $d_1 < d_2$ or $d_1 = d_2$.
In the former case, $\mathcal{N}_{\mathrm{aut}}=2(d_2 - d_1 + 2)$, whereas $\mathcal{N}_{\mathrm{aut}} = 6$ when $d_1 = d_2$.
Combining these informations with Eq. (\ref{basic_counting}), we get, in case $d+2 \leq N$:
\begin{eqnarray}
\mathcal{N}_0 & = & 4 d_1 - 1 \;\;\; (d_1 < d_2) \\
\mathcal{N}_0 & = & 4 d_1 - 3 \;\;\; (d_1 = d_2)
\end{eqnarray}
When $d+2 > N$, and $n=d+2-N$, this value for $\mathcal{N}_0$ has to be subtracted by $n^2$.
These expressions show that, for given $N$ and $d$, the maximal number of $SU(N)$-invariant deformation modes
is obtained by maximizing $d_1$, with the constraint $2d_1 \leq d$. This gives $d_1 = d_2 = d/2$ for even $d$, and
$d_1  = (d-1)/2$, $d_2 = (d+1)/2$ for odd $d$. A table of values of $\mathcal{N}_0$ given by this counting argument,
for $M=2$, $3 \leq N \leq 10$ and $1 \leq d \leq 10$ reads:

\begin{center}
\begin{tabular}{| c || c | c | c | c | c | c | c | c | c | c |}
\hline 
$N$ \textbackslash $d$ & 1 & 2 & 3 & 4 & 5 & 6 & 7 & 8 & 9 & 10 \\ \hline \hline
3 & $\emptyset$ & 0 & $\emptyset$ & $\emptyset$ & $\emptyset$ & $\emptyset$ & $\emptyset$ & $\emptyset$ & $\emptyset$ & $\emptyset$ \\ \hline
4 & $\emptyset$ & 1 & 2 & 1 & $\emptyset$ & $\emptyset$ & $\emptyset$ & $\emptyset$ & $\emptyset$ & $\emptyset$ \\ \hline
5 & $\emptyset$ & 1 & 3 & 4 & 3 & 0 & $\emptyset$ & $\emptyset$ & $\emptyset$ & $\emptyset$ \\ \hline
6 & $\emptyset$ & 1 & 3 & 5 & 6 & 5 & 2 & $\emptyset$ & $\emptyset$ & $\emptyset$  \\ \hline
7 & $\emptyset$ & 1 & 3 & 5 & 7 & 8 & 7 & 4 & $\emptyset$ & $\emptyset$ \\ \hline
8 & $\emptyset$ & 1 & 3 & 5 & 7 & 9 & 10 & 9 & 6 & 1 \\ \hline
9 & $\emptyset$ & 1 & 3 & 5 & 7 & 9 & 11 & 12 & 11 & 8 \\ \hline
10 & $\emptyset$ & 1 & 3 & 5 & 7 & 9 & 11 & 13 & 14 & 13 \\ \hline
\end{tabular}
\end{center}
In this table, the $\emptyset$ sign is used to denote negative values of $\mathcal{N}_0$, which correspond to an absence of $Gr(2,N)$
textures with a uniform topological density. $\mathcal{N}_0=0$ suggests the existence of uniform textures forming a discrete set of $SU(N)$ orbits. 

We see two clear trends in this table. First, increasing $N$ at fixed $d$ increases $\mathcal{N}_0$ until it saturates for $N \geq d+2$. At fixed $N$,
which is probably more relevant to real physical systems, $\mathcal{N}_0$ first increases as $d$ increases, then reaches a maximum and decreases
again. It becomes negative when $d$ becomes sufficiently large. This is reminiscent of the $M=1$ case (projective textures), for which we have seen that 
$\mathcal{N}_0$ is negative as soon as $d \geq N$.  The new feature for $M \geq$ is the prediction of new $SU(N)$-invariant deformation modes
for uniform textures, which have no counterpart for the $M=1$ case.

We should emphasize again that results shown in the above table are conjectural since we have relied on two unproven assumptions. 
It appears that some entries in the table are not correct. For example, let us consider $N=3$ and $M=2$. Sending $M$-dimensional
subspaces of $\mathbb{C}^N$ in to $N-M$ dimensional ones by duality shows that $Gr(2,3)$ is the same manifold as $Gr(1,3)\simeq \mathbb{C}P(2)$.
For $\mathbb{C}P(2)$ textures, $\mathcal{N}_0=0$ for $d=1$ and $d=2$, and there are no uniform solutions for $d \geq 3$. We get the same
pattern as for the $N=3$ row in the above table, excepted for $d=1$. This shows that it is important to test these simple but non rigorous counting arguments
with explicit constructions of uniform solutions for some tractable cases at sufficiently small $d$.


\begin{thebibliography}{99}
\bibitem{odd} In the case of odd $M$ one has to take an $M$-sheeted cover of the principal torus, which leads to a different set of theta-functions.
\end{thebibliography}
\end{document}